\documentclass[tighten,twocolumn]{aastex63}

\usepackage{graphicx}
\usepackage{ulem}
\usepackage{amsmath}
\usepackage{wrapfig}
\usepackage[thinlines]{easytable}
\usepackage{tikz}
\usepackage{hyperref}

\newcolumntype{P}[1]{>{\centering\arraybackslash}p{#1}}

\begin{document}
\shorttitle{}
\shortauthors{}

\title{Hydrodynamics of Collisions and Close Encounters between Stellar Black Holes and Main-sequence Stars}

\correspondingauthor{Kyle Kremer}
\email{kkremer@caltech.edu}

\author[0000-0002-4086-3180]{Kyle Kremer}
\altaffiliation{NSF Astronomy \& Astrophysics Postdoctoral Fellow}
\affiliation{TAPIR, California Institute of Technology, Pasadena, CA 91125, USA}
\affiliation{The Observatories of the Carnegie Institution for Science, Pasadena, CA 91101, USA}

\author[0000-0002-7444-7599]{James C.\ Lombardi Jr.}
\affiliation{Department of Physics, Allegheny College, Meadville, Pennsylvania 16335, USA}

\author[0000-0002-1568-7461]{Wenbin Lu}
\affiliation{Department of Astrophysical Sciences, Princeton University, Princeton, NJ 08544, USA}
\affiliation{Departments of Astronomy, Theoretical Astrophysics Center, UC Berkeley, Berkeley, CA 94720, USA}

\author[0000-0001-6806-0673]{Anthony L. Piro}
\affiliation{The Observatories of the Carnegie Institution for Science, Pasadena, CA 91101, USA}

\author[0000-0002-7132-418X]{Frederic A.\ Rasio}
\affiliation{Center for Interdisciplinary Exploration \& Research in Astrophysics (CIERA) and Department of Physics \& Astronomy \\ Northwestern University, Evanston, IL 60208, USA}

\begin{abstract}

Recent analyses have shown that close encounters between stars and stellar black holes occur frequently in dense star clusters. Depending upon the distance at closest approach, these interactions can lead to dissipating encounters such as tidal captures and disruptions, or direct physical collisions, all of which may be accompanied by bright electromagnetic transients. In this study, we perform a wide range of hydrodynamic simulations of close encounters between black holes and main-sequence stars that collectively cover the parameter space of interest, and we identify and classify the various possible outcomes. In the case of nearly head-on collisions, the star is completely disrupted with roughly half of the stellar material becoming bound to the black hole. For more distant encounters near the classical tidal-disruption radius, the star is only partially disrupted on the first pericenter passage. Depending upon the interaction details, the partially disrupted stellar remnant may be tidally captured by the black hole or become unbound (in some cases, receiving a sufficiently large impulsive kick from asymmetric mass loss to be ejected from its host cluster). In the former case, the star will undergo additional pericenter passages before ultimately being disrupted fully. Based on the properties of the material bound to the black hole at the end of our simulations (in particular, the total bound mass and angular momentum), we comment upon the expected accretion process and associated electromagnetic signatures that are likely to result.
\vspace{1cm}
\end{abstract}


\section{Introduction}

A wide variety of observational \citep[e.g.,][]{Strader2012,Chomiuk2013,Miller-Jones2015,Giesers2018} and computational/numerical evidence \citep[e.g.,][]{Mackey2007,BreenHeggie2013,Morscher2015,Wang2016,Weatherford2020,Kremer2020} suggests that dense star clusters of all masses, from low-mass open clusters to globular clusters and galactic nuclei, may contain significant numbers of stellar-mass black holes throughout their lifetimes. These black holes interact dynamically with one another \citep[e.g.,][]{Rodriguez2016a} and with luminous stars \citep[e.g.,][]{Kremer2018a} in their host cluster. Indeed, the detection of a growing number of black hole binaries with luminous star companions in both old globular clusters \citep[e.g.,][]{Giesers2019} and young star clusters \citep{Saracino2021} highlight the occurrence of such interactions.

As an inevitable consequence of this dynamical mixing, cluster stars occasionally undergo close encounters with black holes. For sufficiently close passages, tides are raised on the star by the black hole that may dissipate enough energy to bind the two objects or disrupt the structure of the star.
The possibility of tidal interactions of stars with compact remnants has been noted in several astrophysical contexts since they were first proposed several decades ago \citep{Fabian1975,Teukolsky_1977,LeeOstriker1986}. 
Depending upon the distance of closest approach as well as the stellar structural details, these close encounters may take the form of physical collisions \citep[e.g.,][]{FryerWoosley1998,ZhangFryer2001}, tidal disruptions \citep[e.g.,][]{Perets_2016,Kremer2019c}, tidal captures \citep[e.g.,][]{Fabian1975,Ivanova2017}, or more distant tidal encounters that only weakly perturb the star \citep[e.g.,][]{AlexanderKumar2001}.

If during the encounter material is stripped from the star and becomes bound to the black hole, a fraction of the stellar material may be accreted by the black hole, producing a luminous flare. Previous studies \citep[e.g.,][]{Perets_2016,Kremer2019c,Kremer2021a} have demonstrated that in the tidal disruption regime, bright X-ray transients with optical/ultraviolet counterparts likely result. The peak optical luminosities and rates of these tidal disruption events (TDEs) make them notable sources for optical transient observatories such as the Vera Rubin Observatory (VRO) and the Zwicky Transient Facility (ZTF). 
Indeed, transients already detected such as fast blue optical transients like AT2018cow \citep{Margutti2019} may be explained by such TDEs \citep{Kremer2021a}. Furthermore, a number of recent observations have linked TDEs of white dwarfs by massive black holes with massive star clusters \citep[e.g.,][]{Krolik2011,Jonker2013,2018NatAs...2..656L}

Aside from their intrinsic interest as electromagnetic transient sources, close encounters between black holes and stars may have an important effect on the overall demographics of black holes in clusters. If a significant fraction of the disrupted stellar material is accreted by the black hole, these TDEs may lead to significant black hole mass growth, potentially enabling formation of black holes within or above the pair-instability mass gap \citep[e.g.,][]{Giersz2015,Rizzuto2021,ArcaSedda2021,Rose2022}. Furthermore, if a star is disrupted by a binary black hole, the disrupted material may alter the spin magnitude and orientation of the binary black hole components \citep[e.g.,][]{Lopez2019}. In any of these scenarios, these TDEs would have important implications upon the properties of binary black hole mergers occurring in dense star clusters that may be detectable as gravitational wave sources by LIGO/Virgo and future detectors like LISA \citep[e.g.,][]{Kremer2019b}.

Studies of the hydrodynamics of close encounters between black holes and stars are critical to elucidate these various possible outcomes and their details. In contrast to the supermassive black hole TDE regime \citep[for recent reviews see][]{Stone2019, Gezari2021}, the hydrodynamics of TDEs in the stellar-mass black hole regime have been relatively unexplored, with a few exceptions that focused on distant encounters near the classical tidal disruption radius \citep[e.g,][]{Perets_2016,Lopez2019,Wang2021}. The direct collision regime where the pericenter distance is comparable to or less than the stellar radius has not been explored at all in this context. In this paper, we investigate this topic using the smoothed particle hydrodynamics (SPH) code \texttt{StarSmasher} \citep[e.g.,][]{RasioThesis,Gaburov_GPU_SPH} which was designed specifically to compute stellar collisions and tidal interactions \citep[e.g.,][]{Lombardi2006, Ivanova2010,Antonini2011,Hwang2017}. We focus in particular on interactions of \textit{main-sequence} stars with stellar black holes, which are expected to be the most common type of black hole--star encounter \citep{Kremer2019c}. The disruptions of giant branch stars \citep[e.g.,][]{Ivanova2017} and white dwarfs \citep[e.g.,][]{Rosswog2009} by stellar black holes are of high interest in their own right and likely produce markedly different outcomes from the main-sequence star case. Therefore we reserve analysis of black hole encounters with white dwarfs and giants for future study.

We compute SPH simulations for a wide range of black hole -- main-sequence star encounters. Specifically, we vary the black hole mass (mostly $10\,M_{\odot}$ with a subset of $30\,M_{\odot}$ models), the mass of the disrupted star ($0.5-20\,M_{\odot}$), the initial structure of the disrupted star (represented by its polytropic index), and the distance of closest approach (ranging from perfectly head-on collisions to widely off-axis interactions with pericenter distances several times the classical tidal disruption limit). Altogether, our suite of SPH simulations spans nearly completely the parameter space predicted in recent $N$-body simulations for tidal encounters between main-sequence stars and black holes in realistic star clusters \citep[e.g.,][]{Kremer2019c,Kremer2020}. This suite of models is a critical tool toward understanding the basic hydrodynamic outcomes of these encounters, which set the stage for more detailed future simulations including additional effects such as radiation, accretion feedback, magnetic fields, etc., and for the inclusion of more detailed treatments of these events within $N$-body cluster models.

The subsequent sections of this paper are organized as follows. In Section~\ref{sec:methods}, we discuss our numerical methodology and the types of analysis used in the SPH simulations. In Section~\ref{sec:results}, we present our simulation results and describe distinct hydrodynamic outcomes across the parameter space. In Section~\ref{sec:EM} we discuss possible electromagnetic signatures that may result, based on the features of our SPH models. In Section~\ref{sec:rates}, we discuss the relative event rates of these various outcomes expected in realistic clusters. Finally, we summarize and discuss our results in Section~\ref{sec:summary}.

\section{Methods}

\label{sec:methods}

To perform the hydrodynamic calculations, we use the code \texttt{StarSmasher} (\citealp{Evghenii_2018}; originally \citealp{RasioThesis}) to treat the hydrodynamics. Like other SPH codes \citep[e.g.,][]{MONAGHAN1997298,Springel2005}, \texttt{StarSmasher} employs a Lagrangian method which represents the stellar gas as a collection of particles, each with an extended density profile given by a smoothing kernel function. 
For these calculations, we use a Wendland C4 kernel \citep{Wendland1995} with compact support $2h$ (where $h$ is the smoothing length) for smoothing and gravitational softening.  The artificial viscosity prescription is coupled with a Balsara switch \citep{Balsara1995} to prevent unphysical inter-particle penetration, implemented as described in \citet{2015ApJ...806..135H}. Gravitational forces and energies between particles are computed directly using direct summation on NVIDIA graphics cards as described in \citet{Gaburov_GPU_SPH} and in accord with the energy-conserving formalism of \citet{2007MNRAS.374.1347P}.
Using direct summation for gravitational force calculations instead of a tree-based method has been shown to yield more accurate results at the cost of speed \citep{Gaburov_GPU}. 
All of the hydrodynamic calculations are computed using an analytic equation of state incorporating ideal gas and radiation pressure, implemented as in \citet{Lombardi2006}.

Polytropic density profiles are used to generate all initial stellar models. For low-mass stars ($M_{\star}=0.5\,M_{\odot}$), we use polytropic models with $n=1.5$ and $\Gamma=5/3$, appropriate for fully or deeply convective M-dwarf stars. For more massive stars ($M_{\star}\geq 1\,M_{\odot}$), which have no, or relatively shallow, convective envelopes, we use an Eddington standard model, namely a $n=3$ density profile with an uncoupled adiabatic index $\Gamma$ calculated as in \citet{Rasio1993_Gamma}. The constant ratio of gas to radiation pressure in the Eddington standard model is primarily dependent on the mass of the star, but there is a slight additional dependence on metallicity.  Qualitatively, this dependence arises because changing the metallicity changes slightly the mean molecular weight and therefore the gas pressure if everything else were to remain unchanged. Here, all stellar models were created using the metallicity $Z = 0.1 Z_\odot$ \citep[appropriate for typical old globular clusters; e.g.,][]{Harris1996}. Our stellar models are chemically homogeneous so we are implicitly assuming they are representative of \textit{zero-age} main-sequence stars. Nuances associated with the distinction in TDE dynamics arising from different stellar ages (on the main sequence) are beyond the scope of this initial study.\footnote{For discussion of the effect of stellar age on TDEs in the supermassive black hole regime, see for example \citet{GallegosGarcia2018}.} Finally, we also assume all models are initially nonrotating. The parameters of our stellar models are summarized in Table \ref{table:parents}.

\begin{deluxetable}{ccccc}
\tabletypesize{\footnotesize}
\tablecaption{Initial properties of stellar models \label{table:parents}}
\tablehead{
	\colhead{\hspace{.5cm}$^1M_{\star,i}$}\hspace{.5cm} &
	\colhead{\hspace{.5cm}$^2R_{\star,i}$}\hspace{.5cm} &
	\colhead{\hspace{.5cm}$^3\Gamma$}\hspace{.5cm}&
	\colhead{\hspace{.5cm}$^4n$}\hspace{.5cm}&
	\colhead{\hspace{.5cm}$^5N$}\hspace{.5cm}
}
\startdata
0.5 & 0.7 & 5/3 & 1.5 & $1.57\times 10^5$\\
1 & 1.0 & 1.6660 & 3 & $1.40\times 10^5$ \\
2 & 1.5 & 1.6641 & 3 & $1.33\times 10^5$ \\
5 & 2.6 & 1.6516 & 3 & $1.16\times 10^5$\\
10 & 4.0 & 1.6192 & 3 & $1.25\times 10^5$ \\
15 & 5.1 & 1.5869 & 3 & $1.18\times 10^5$ \\
20 & 6.0 & 1.5600 & 3 & $1.09\times 10^5$
\enddata
\tablecomments{\footnotesize Initial properties of all main-sequence stars used in our SPH simulations. In columns 1-2, we give star mass and radius in solar units.  In column 3, we give the adiabatic index and in column 4, the polytrope index. Finally, in column 5, we give the number of SPH particles employed. As described in the text, for the $M_{\star}=0.5\,M_{\odot}$ case, $\Gamma$ and $n$ are coupled (using standard $\Gamma=1/n+1$ relation). For $M_{\star}\geq1\,M_{\odot}$ cases, we use Eddington standard models where $n$ and $\Gamma$ are uncoupled as calculated in \citet{Rasio1993_Gamma}.}
\end{deluxetable}

The black holes in the hydrodynamic calculations are treated as point particles with a softened potential, and only interact via gravity with the gas SPH particles. In particular, we soften the gravity of the black hole according to the mass distribution of the Wendland C4 kernel with compact support $2h_{\rm BH}$, with $h_{\rm BH} = 0.05$, calculated as described by \citet{2007MNRAS.374.1347P}.  Consequently, the gravitational potential of the black hole approaches a finite value in the limit of zero separation: specifically it approaches $-55/32\,GM_{\rm BH}/h_{\rm BH}$ where $M_{\rm BH}$ is the black hole mass. Our calculations do not employ any accretion or black hole feedback mechanisms. We discuss in more detail the possible role of accretion feedback in Section \ref{sec:future}.

Each simulation employs about $N=1.1\times 10^5$ to $1.6\times 10^5$ equal-mass SPH particles. All of the simulations are run until at least the first pericenter passage was finished (i.e., when the star is either visually fully disrupted or the simulation reached a slowly-varying state such as tidal filaments). In a select few cases, we simulate multiple pericenter passages until the star is fully disrupted.

\subsection{Smoothing Length Determination}

In SPH, it is useful to dynamically adjust smoothing lengths in order to always and everywhere maintain reasonable resolution and neighbor numbers: consequently, SPH particles in compressed regions will obtain smaller smoothing lengths.  
Unless handled carefully, these smoothing lengths can become exceedingly small for material in the vicinity of a black hole and drive the timestep toward zero, effectively halting the simulation.  
In addition, ejected particles can have their smoothing lengths increased to undesirably large values, comparable to their separation from the black hole.  
To avoid such issues, the analytic relation we use for the smoothing length enforces minimum and maximum possible values, namely $h_{\rm f}$ and $h_{\rm c}$ respectively (here the subscript ``f'' stands for floor, while the subscript ``c'' stands for ceiling). 
In particular, the smoothing length of particle $i$ is given by
\begin{equation}
    h_i = \left( \frac{1}{h_{\rm c} - h_{\rm f}} + b \rho_i^{1/3}\right)^{-1} + h_{\rm f}, \label{h}
\end{equation}
where the density $\rho_i$ is
\begin{equation}
    \rho_i = \sum_j m_j W({\bf r}_i-{\bf r}_j, h_i) \label{rho}
\end{equation}
and $W$ is the SPH kernel.  Note that in the small and large density limits, the smoothing length approaches $h_{\rm c}$ and $h_{\rm f}$ respectively.  In this paper, we implement $h_{\rm f}=0.015$ and $h_{\rm c}=100$.  We choose the constant $b$ for each model such that $b (M_{*,i}/N)^{1/3}$ is about 0.8 to 0.9, which yields $\sim 10^2$ neighbors in the outer portion of the parent star. Throughout the simulation, we then simultaneously solve Equations (\ref{h}) and (\ref{rho}) for each particle $i$ at each timestep.  Equation~(\ref{h}) generalizes the analogous equation used in \cite{Lombardi2006} by introducing the floor smoothing length $h_{\rm f}$.  While the use of non-zero $h_{\rm f}$ can sacrifice spatial resolution, it allows the code to run stably and avoids vanishingly small timesteps even when gas becomes highly compressed in the immediate vicinity of the black hole.

We integrate the SPH equations of motion using a leapfrog scheme with shared timesteps calculated by
\begin{equation}
  \Delta t={\rm min}(\Delta t_{\rm gas-gas}, \Delta t_{\rm gas-BH}).
\end{equation}
The $\Delta t_{\rm gas-gas}$ term considers interactions among SPH particles and is determined as described in Appendix~A of \citet{Gaburov_GPU_SPH} with $C_{N,1}=0.1$ and
$C_{N,2}=0.01$.  The $\Delta t_{\rm gas-BH}$ term accounts for interactions between SPH particles and the black hole and is calculated according to
\begin{equation}
  \Delta t_{\rm gas-BH}={\rm Min}_i \left[\left(\Delta t_{{\rm v},i}^{-1}+ \Delta
    t_{{\rm a},i}^{-1} \right)^{-1}\right]\,.
  \label{good.dt}
\end{equation}
For each SPH particle $i$, we use
\begin{equation}
  \Delta t_{{\rm v},i}=C_{N,{\rm v}}
         \frac{r_{{\rm s},i}}{v_{{\rm BH},i}}
         \label{dtv}
\end{equation}
with $r_{{\rm s},i}\equiv \left( r_{{\rm BH},i}^2 + h_{\rm BH}^2 \right)^{1/2}$
being a ``softened separation'' of particle $i$ and the black hole, $r_{{\rm BH},i}$ being their actual separation, and $v_{{\rm BH},i}$
being their relative speed.  Here
\begin{equation}
  \Delta t_{{\rm a},i}=C_{N,{\rm a}}
  \left(\frac{r_{{\rm s},i}}{a_{{\rm BH},i}} \right)^{1/2},
  \label{dta}
\end{equation}
where $a_{{\rm BH},i}$ is the relative acceleration of the SPH particle $i$ to the black hole.
In this paper we use $C_{N,{\rm v}}=C_{N,{\rm a}}=0.01$ and we soften the gravity of the black hole according to the mass distribution of the SPH kernel, with smoothing length $h_{\rm BH} = 0.05$. Our time-stepping allows for total energy conservation to within 0.004\% in the majority of the simulations of this paper and always to within 0.5\%.  Total angular momentum is conserved to within 0.0003\% in the majority of our simulations and always to within $10^{-4} (GM_\odot^3R_\odot)^{1/2}$ on an absolute scale.

\subsection{Analysis of Hydrodynamics}

For each output snapshot from the hydrodynamic calculations, we determine the mass $M_{\star}$ bound to the star (component $j=1$) and the mass $M_{\rm BH}$ of the black hole plus the mass bound to it (component $j=2$) using an iterative procedure similar to the one in \cite{Lombardi2006}. To be bound to one of these two components, an SPH particle must, at a minimum, have a negative total energy with respect to the center of that component.  For the center of the star, we choose the location of its highest density particle, while for the center of the black hole and its bound mass, we choose the black hole particle itself.  More precisely, for a particle to be considered bound to component $j$, the quantity $v_{ij}^2/2 - GM_j/d_{ij} + u_i , $ must be negative, where $v_{ij}$ is the velocity of particle $i$ with respect to the center of component $j$, $d_{ij}$ is the distance from particle $i$ to that center, and $u_i$ is the specific internal energy of particle $i$. If this quantity is negative for both $j = 1$ and $j=2$, then the particle is assigned to the component $j$ that has a closer center.  All mass not considered bound to either component is included in the mass $M_{\rm ej}$ of the ejecta.
Once the particles have been sorted among the components, we compute several properties of the mass bound to the black hole, as discussed in Section \ref{sec:results}.

\subsection{Choice of simulation grid parameters}

\begin{figure}
    \centering
    \includegraphics[width=\columnwidth]{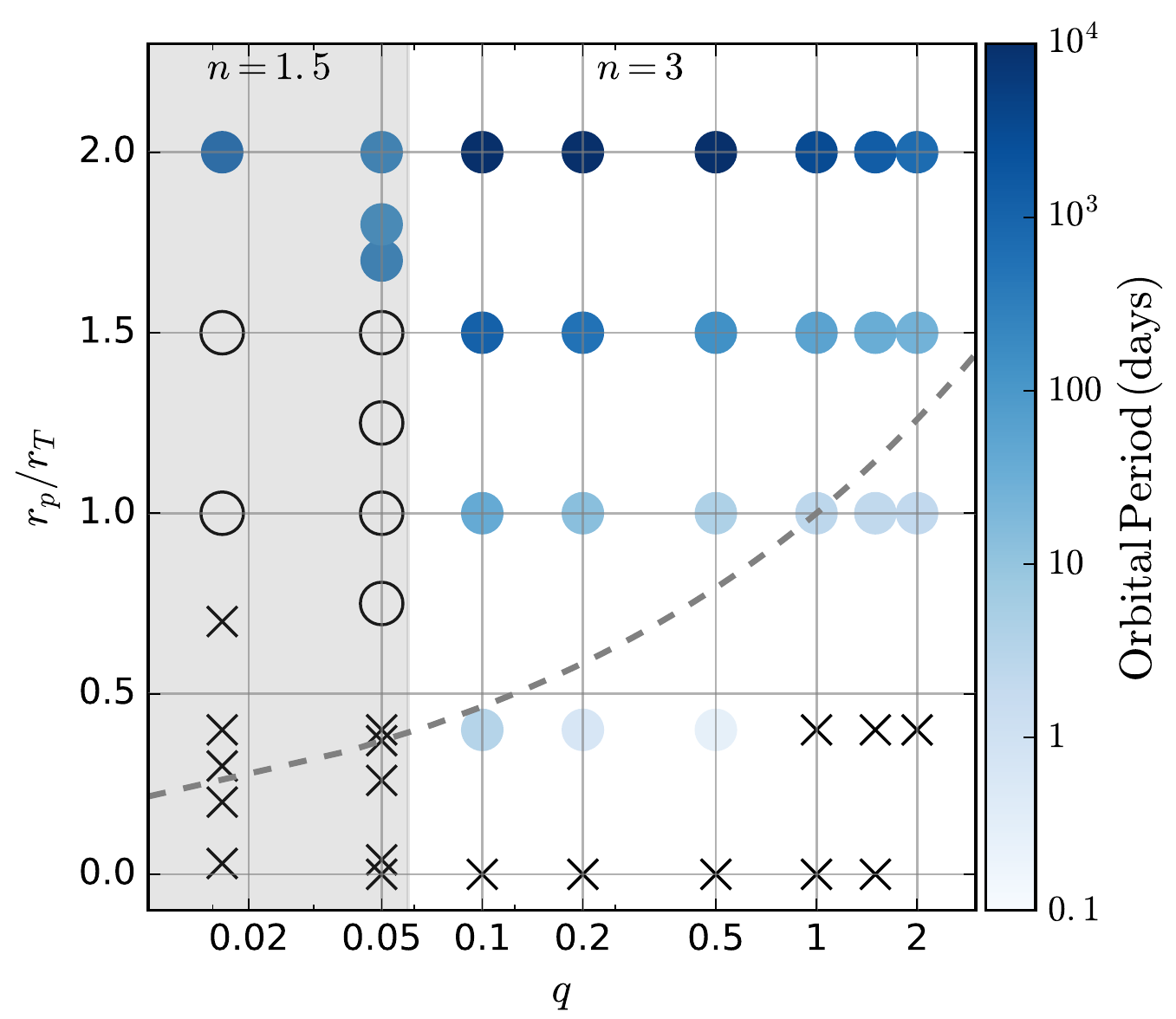}
    \caption{\footnotesize Summary of outcomes of the first pericenter passage for all SPH simulations computed in this study. On the horizontal axis, we show the mass ratio, $q=M_{\star}/M_{\rm{BH}}$. With the exception of the $q=0.016$ case (where we assume $M_{\rm{BH}}=30\,M_{\odot}$), all other simulations shown in this figure adopt $M_{\rm{BH}}=10\,M_{\odot}$ and only the stellar mass is varied. Black X's denote models where the star is disrupted completely by the black hole on the first passage. Blue filled circles denote models where the star is partially disrupted and the stripped stellar remnant is tidal captured by the black hole, forming a bound binary (with the blue shade denoting the binary orbital period). Open circles denote models where the star is partially disrupted but remains unbound, resulting from an impulsive kick from asymmetric mass loss at pericenter. The dashed gray line shows the boundary for which $r_p<R_{\star}$ for a given mass ratio. Finally, the gray shaded background denotes models for which an $n=1.5$ polytropic index was used. All other models adopted an $n=3$ index.}
    \label{fig:summary}
\end{figure}

We run 49 independent simulations in this study varying the masses of the star and black hole, the pericenter distance of the encounter, and the polytropic index of the star. We assume primarily a black hole mass of $10\,M_{\odot}$ and additionally run a small subset of models that adopt a black hole mass of $30\,M_{\odot}$ (with the exception of Figure \ref{fig:summary}, all figures in the paper show results exclusively for the $M_{\rm{BH}}=10\,M_{\odot}$ case). These choices are motivated by recent N-body studies of black hole TDEs in star clusters \citep{Kremer2019c,Kremer2021a}. We assume a wide range in stellar masses from $0.5-20\,M_{\odot}$, which reasonably trace the range of the initial mass function expected for clusters at birth. 

For each set of black hole+star masses, we perform simulations for a range of pericenter distances. A characteristic length scale for close tidal encounters is the classical tidal disruption radius of the star

\begin{equation}
    \label{eq:r_TD}
    r_T = \Big(\frac{M_{\rm{BH}}}{M_{\star}} \Big)^{1/3} R_{\star}.
\end{equation}
where $M_{\rm{BH}}$ is the black hole mass, $M_{\star}$ is the stellar mass, and $R_{\star}$ is the stellar radius.
We perform models with pericenter distances ranging from $r_p=0$ (directly head-on collisions) to $r_p = 2r_T$ (distant tidal encounters where negligible mass is stripped from the star).

As discussed in \citet{Kremer2019c}, in typical clusters tidal disruptions of low-mass main-sequence stars ($M_{\star}\lesssim1\,M_{\odot}$) are much more common than those of high-mass stars. Low-mass stars are more numerous for a standard initial stellar mass function \citep[e.g.,][]{Kroupa2001} and also have much longer main-sequence lifetimes compared to higher mass stars. The median mass of stars that undergo TDEs in \citet{Kremer2019c} was found to be roughly $0.5\,M_{\odot}$. In order to investigate further this more frequent case, we perform additional simulations (a finer grid of pericenter distances) of the $M_{\star}=0.5\,M_{\odot}$, $M_{\rm{BH}}=10\,M_{\odot}$ combination.

Since typical energy scales within a globular cluster are close to zero \citep[e.g.,][]{HeggieHut2003}, most interactions occur in the gravitational focusing regime, with low values of $v_\infty$. 
For this reason, we simulate strictly parabolic orbits with $v_\infty = 0$.  We work in the center-of-mass frame of the system, with the orbit in the $xy$ plane. Future studies may examine the dynamics of encounters in the hyperbolic regime, especially the case of $v_\infty > 100\,\rm{km/s}$ which is relevant for relatively massive super star clusters and nuclear star clusters.

Table \ref{table:sims} contains the full list of simulations run for this study. Columns 1-4 list the initial conditions for each simulation and the remaining columns contain information on the outcomes of the simulations as described in subsequent sections.

\bigskip

\section{Results of Hydrodynamic models}
\label{sec:results}

\begin{figure*}
    \centering
    \includegraphics[width=\linewidth]{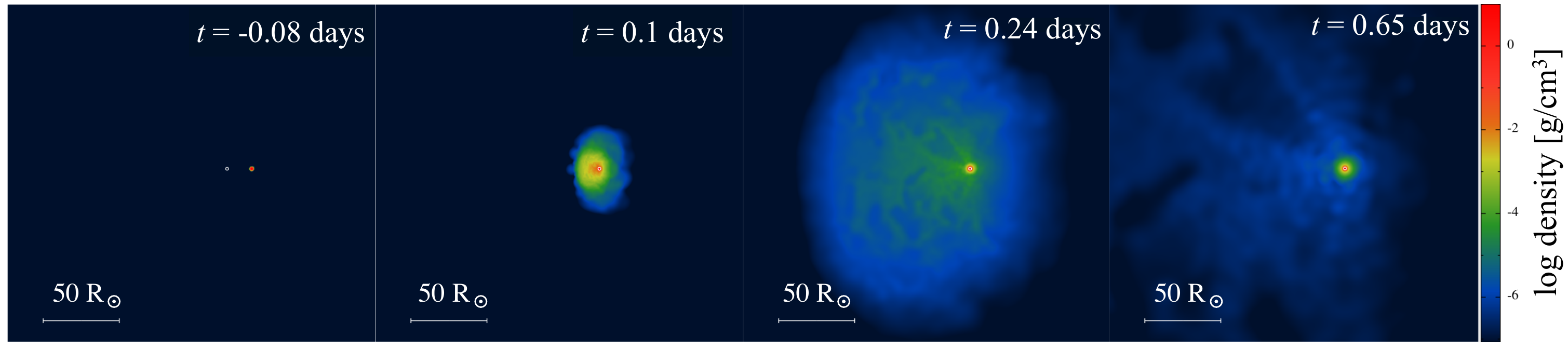}
    \caption{\footnotesize A head-on collision ($r_p=0$) between a $2\,M_{\odot}$ star (Eddington standard model) and a $10\,M_{\odot}$ black hole. The star is destroyed completely through the black hole encounter. After the encounter, roughly half of the disrupted material is unbound from the system and half remains bound to the black hole. In response to this mass loss, the black hole receives an impulsive kick of roughly $100\,\rm{km\,s}^{-1}$ as described in Section \ref{sec:COM_kick}. For video of this simulation \href{https://sites.northwestern.edu/kremerastronomy/files/2021/12/2_10.mov}{click here}.
    }
    \label{fig:full_disrupt}
\end{figure*}

\begin{figure*}
    \centering
    \includegraphics[width=\linewidth]{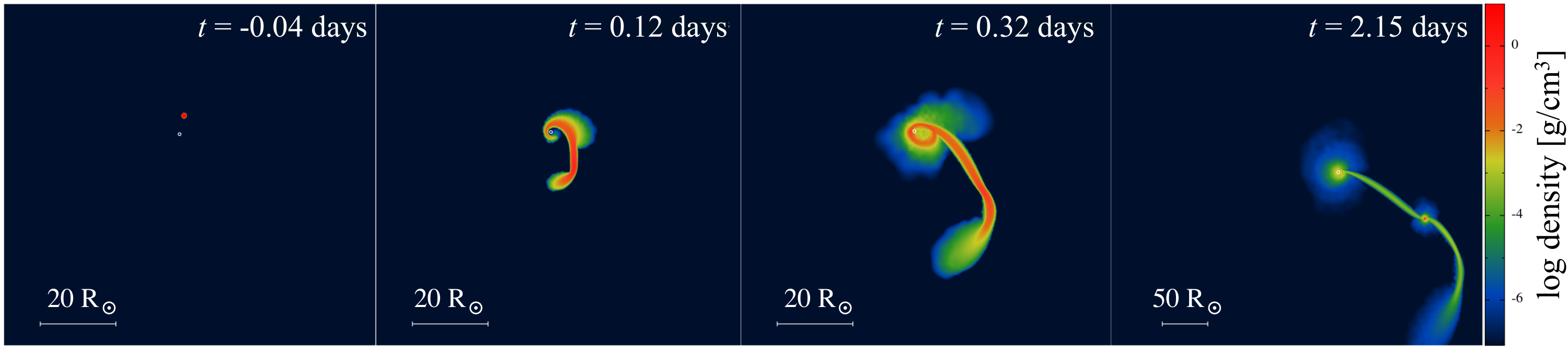}
    \caption{\footnotesize A $0.5\,M_{\odot}$ star (modelled as an $n=1.5$ polytrope) encountering a $10\,M_{\odot}$ black hole with $r_p=r_T$. The star is partially disrupted on first passage and receives an impulsive kick from asymmetric mass loss at pericenter. The roughly $0.2\,M_{\odot}$ partially-disrupted stellar remnant has final velocity of roughly $260\,\rm{km\,s}^{-1}$ as described in Section \ref{sec:unbound_vs_bound}. For a video of this simulation \href{https://sites.northwestern.edu/kremerastronomy/files/2021/12/05_10.mov}{click here}.}
    \label{fig:partial_unbound}
\end{figure*}

\begin{figure*}
    \centering
    \includegraphics[width=0.8\linewidth]{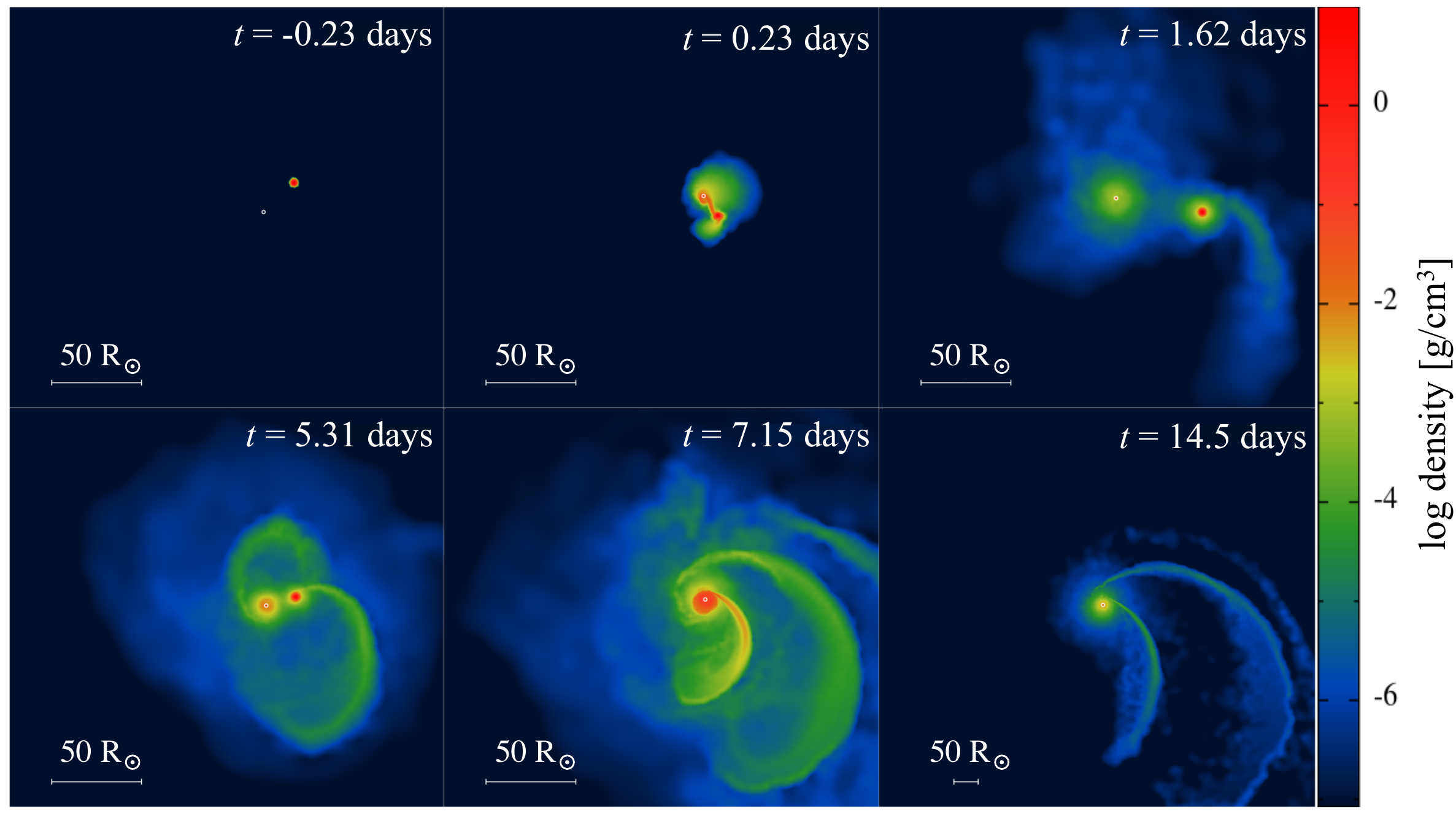}
    \caption{\footnotesize A $5\,M_{\odot}$ star (Eddington standard model) encountering a $10\,M_{\odot}$ black hole with $r_p=r_T$. The star is partially disrupted on first passage and tidally captured by the black hole. The partially disrupted remnant returns to pericenter for a second time roughly 4 days after the first passage and undergoes 4 total passages before ultimately being disrupted completely by the black hole (see Figure \ref{fig:evol_insp}). For a video of this simulation \href{https://sites.northwestern.edu/kremerastronomy/files/2021/12/5_10.mov}{click here}.
    }
    \label{fig:partial_bound}
\end{figure*}

In this section, we describe the results of our suite of hydrodynamic models. In Figure \ref{fig:summary}, we summarize the various outcomes of the first pericenter passage for the simulations computed in this study. On the vertical axis, we show the penetration factor of the interaction, $r_p/r_T$. On the horizontal axis, we show the mass ratio, $q=M_{\star}/M_{\rm{BH}}$. With the exception of the $q=0.016$ case on the far left (where we assume $M_{\rm{BH}}=30\,M_{\odot}$; simulations 1-8 in Table \ref{table:sims}), all other simulations shown in this figure adopt $M_{\rm{BH}}=10\,M_{\odot}$ and only the stellar mass is varied. The gray shaded background denotes models for which an $n=1.5$ polytropic index was used (for these models, we adopt $M_{\star}=0.5\,M_{\odot}$). All other models adopted an $n=3$ index (in this case, we adopt $M_{\star}\geq1\,M_{\odot}$). We identify three distinct outcomes, described below. In Figures \ref{fig:full_disrupt}, \ref{fig:partial_unbound}, \ref{fig:partial_bound} we show the hydrodynamic evolution for an example simulation of each of these cases.\footnote{Figures \ref{fig:full_disrupt}, \ref{fig:partial_unbound}, and \ref{fig:partial_bound} and accompanying animations were created using the \texttt{SPLASH} visualization software \citep{Splash2007}.}

\textbf{(i) Complete tidal disruption:} This outcome is indicated by X's in Figure \ref{fig:summary}. Here, the star is completely destroyed by the encounter with the black hole. As expected, the most penetrating interactions (smaller $r_p/r_T$) are most likely to lead to full disruption. In particular, perfectly head-on collisions ($r_p=0$) lead to full disruptions for all mass ratios. At high mass ratios ($q\geq 1$), less penetrating interactions ($r_p/r_T \leq 0.4$) also lead to complete disruption. This is not surprising, since for $q > 1$, the tidal disruption radius of the star is less than its physical radius (see Equation \ref{eq:r_TD}). In this case, a fixed value of $r_p/r_T$ is more penetrating at high mass ratios relative to lower mass ratios, and thus is more likely to lead to full disruption.

At lower mass ratios ($q\leq0.05$), full disruption also occurs for less penetrating encounters. This is a direct result of the different polytropic index. An $n=1.5$ polytrope (these models are marked by the gray background in Figure \ref{fig:summary}) is more uniformly distributed than the $n=3$ case. Thus, a larger amount of mass is stripped by the black hole at pericenter, making it relatively easy to disrupt the star completely than in the more centrally concentrated $n=3$ polytrope case. Indeed, the choice of polytropic index plays a key role in the precise outcome for encounters at larger pericenter distances, as discussed next.

At even lower mass ratios not considered in our models, full disruption is expected to occur at even wider pericenter distances. For example, \citet{Perets_2016} showed that Jupiter-mass objects are disrupted fully by a $10\,M_{\odot}$ black hole for $r_p=r_T$ (also using SPH models computed with \texttt{StarSmasher}). Such mass ratios are more similar to the more well-studied supermassive black hole TDE regime, where full disruptions are, in general, expected \citep[e.g.,][]{Stone2019}.

We show an example of the complete disruption case in Figure \ref{fig:full_disrupt}. Here, we show evolution of a $2\,M_{\odot}$ star ($n=3$ polytrope) interacting with a $10\,M_{\odot}$ black hole with pericenter distance $r_p=0$. At the end of the simulation, roughly half (here $1\,M_{\odot}$) of the disrupted stellar material is bound to the black hole with the remaining half being unbound. For encounters with $q\leq 1$, this result (half bound/half unbound) holds in general for all perfectly head-on collisions.

\textbf{(ii) Partial tidal disruption+unbound remnant:} For encounters with larger $r_p/r_T$, the star is only partially disrupted on the first passage. In this case, a fraction of disrupted stellar material becomes unbound from the system and the remainder is bound to the black hole. Depending on the details of the stellar mass loss (to be discussed in Section \ref{sec:unbound_vs_bound}), the partially disrupted stellar remnant may be either unbound (as discussed here) or bound (as discussed next) to the black hole. The partial disruption+unbound outcome is indicated by open circles in Figure \ref{fig:summary}. Here the partially disrupted stellar remnant receives an impulsive ``kick'' from asymmetric ejecta mass loss and remains unbound following the first passage. We show an example simulation of this case in Figure \ref{fig:partial_unbound} for a $0.5\,M_{\odot}$ star ($n=1.5$ polytrope) interacting with a $10\,M_{\odot}$ black hole with pericenter distance $r_p=r_T=2.7\,R_{\odot}$. In this case, $0.23\,M_{\odot}$ (46\%) of material is bound to the black hole, $0.07\,M_{\odot}$ (14\%) of material is unbound from both the star and the black hole, and the final mass of the partially disrupted remnant is $0.2\,M_{\odot}$. 

We note that the partial disruption outcome identified here is distinct from the situation explored in previous studies where the star is destroyed completely and then later reforms from the disrupted debris \citep[e.g.,][]{Miles2020,Nixon2021,Nixon2022}. In our partial disruption simulations, the stellar core remains intact throughout. In the cases of more penetrating encounters where the star is indeed destroyed fully, we do not consider the possibility of reformation of stars (or other objects) through gravitational instabilities of the disrupted debris on longer timescales. For discussion of this possibility in the context of TDEs, see \citet{Kochanek1994,CoughlinNixon2015,CoughlinNixon2020,Wang2021}.

\textbf{(iii) Partial tidal disruption+tidal capture:} If sufficient orbital energy is injected into the oscillation modes of the star during the close passage, the partially disrupted remnant will become bound to the black hole. This outcome is indicated by filled blue circles in Figure \ref{fig:summary}. The blue shade denotes the orbital period of the new black hole+star binary, which corresponds to the time until the second pericenter passage. We show an example simulation of this case in Figure \ref{fig:partial_bound} for a $5\,M_{\odot}$ star interacting with a $10\,M_{\odot}$ black hole with pericenter distance $r_p=r_T=2.6\,R_{\odot}$.

\begin{figure}
    \centering
    \includegraphics[width=\columnwidth]{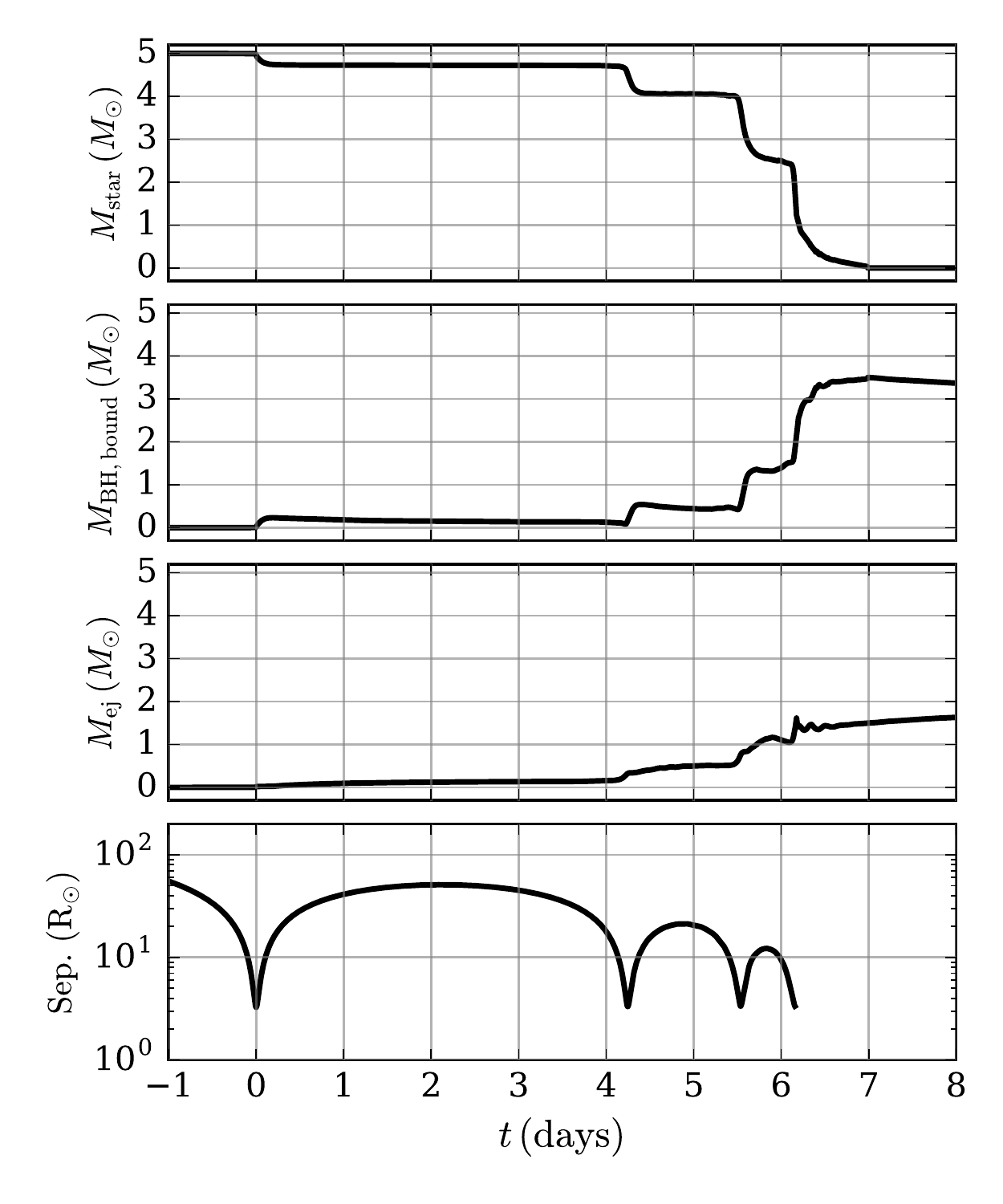}
    \caption{From top to bottom, time evolution of the mass of the star, mass of material bound to the black hole, mass of material unbound from the system, and separation between the black hole and star for the $5\,M_{\odot}$ star + $10\,M_{\odot}$ black hole simulation shown in Figure \ref{fig:partial_bound}. After each pericenter passage, a small amount of material is stripped from the star until, ultimately, the star inspirals completely and is fully disrupted by the black hole.}
    \label{fig:evol_insp}
\end{figure}

After the first passage, $0.22\,M_{\odot}$ of stripped material (roughly $5\%$ of initial stellar mass) becomes bound to the black hole, $0.05\,M_{\odot}$ of material ($1\%$ of total) is unbound from the system, and the final mass of the partially disrupted remnant is $4.73\,M_{\odot}$. The time until the second pericenter passage is just over $4\,$days. In total, this system undergoes 4 pericenter passages before the star is fully disrupted by the black hole. At the end of the simulation (after full disruption of the star), roughly $3\,M_{\odot}$ of material is bound to the black hole and roughly $2\,M_{\odot}$ has been unbound from the system. We expect inspiral and full disruption is the eventual outcome of all instances where the star is partially disrupted and tidally captured by the black hole after the first passages, with the exact number of passages determined by the mass ratio and pericenter distance (we discuss this point in further detail in Section \ref{sec:multiple_passage}). In Figure \ref{fig:evol_insp}, we show, from top to bottom, the time evolution of the total stellar mass, the mass of material bound to the black hole, the mass of material unbound from the system, and the separation between the black hole and star. 

In columns 5, 6, and 7 of Table \ref{table:sims} we show, respectively, the total mass bound to the black hole, total mass remaining (if any) of the disrupted star, and the total mass of material that has been unbound entirely from the system. In column 13 we show the orbital period for the partially disrupted remnant to return to pericenter a second time. Instances where the star is fully disrupted on the first passage or where the disrupted remnant is unbound from the black hole are marked ``N/A.''

With the basic features laid out for the three characteristic outcomes, we now discuss the key dynamical processes at play that determine the details of these distinct outcomes for the various simulations.

\subsection{Tidal capture versus ``hypervelocity'' stellar core}
\label{sec:unbound_vs_bound}

\begin{figure}
    \centering
    \includegraphics[width=0.85\columnwidth]{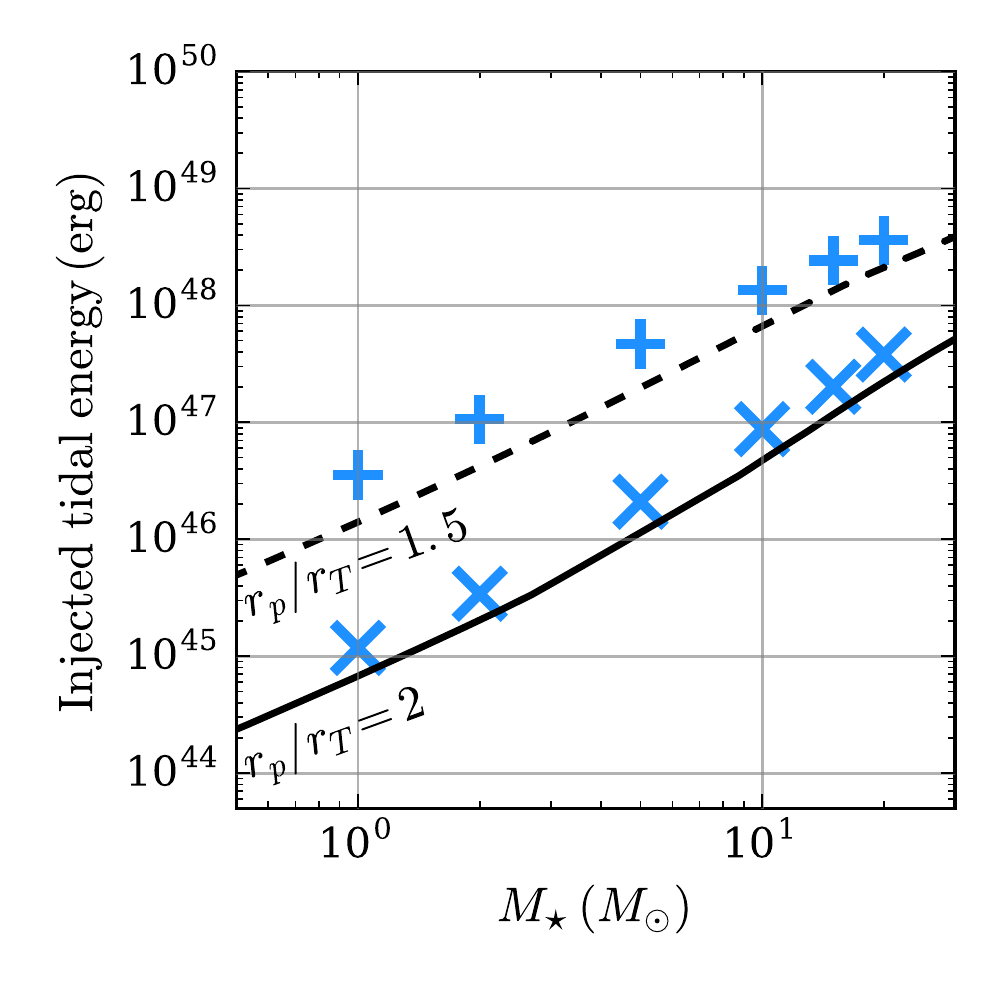}
    \caption{\footnotesize Total energy injected into the star through tides computed for simulations with $r_p/r_T=1.5$ (``+'' symbols) and $r_p/r_T=2$ (``x'' symbols) compared to energy computed from the analytic expression in Equation~(\ref{eq:Etides}). The good agreement between simulations and theory demonstrates the tidal capture process is modeled effectively in our SPH simulations.}
    \label{fig:E_tides}
\end{figure}

A key process that affects the dynamics of these encounters is the injection of orbital energy into oscillation modes of the star through tides. In instances where the star loses no mass, this is a well-known result \citep[e.g.,][]{Teukolsky_1977} and is understood to provide a method to form compact binaries through dynamical encounters in dense star clusters \citep[e.g.,][]{Fabian1975}. The change in orbital energy due to tides for an initially non-oscillating star is

\begin{equation}
    \label{eq:Etides}
    E_{\rm{tides}} \approx -\Big(\frac{M_{\rm{BH}}}{M_\star}\Big)^2 \frac{G M_\star^2}{R_{\star}} \Bigg[ \Big(\frac{R_{\star}}{r_p}\Big)^6 T_2 + \Big(\frac{R_{\star}}{r_p}\Big)^8 T_3 \Bigg]
\end{equation}
where $T_2$ and $T_3$ are dimensionless functions (note we have included here only the first two terms)
that measure energy contribution from different harmonic modes \citep[for typical values of $T_l$ for various polytropic indices, see e.g.,][]{LeeOstriker1986}.

In Figure \ref{fig:E_tides}, we show $\Delta E_{\rm{tides}}$ (computed from Equation \ref{eq:Etides}) versus initial stellar mass for $r_p/r_T=1.5$ (dashed curve) and $=2$ (solid curve). $T_2$ and $T_3$ are taken from Figure 1 of \citet{LeeOstriker1986} for the $n=3$ case. We show, as ``+'' and ``x'' symbols, the total tidal energy injected into the star after the first pericenter passage from our relevant SPH simulations. This energy value is computed simply as the change in total internal energy of the star before and after the first pericenter passage. In the limit shown here where little mass is stripped from the star (see also column 6 of Table \ref{table:sims}), the simulation results agree quite well with the analytic theory, demonstrating that the tidal capture process is effectively modeled in our simulations.

For encounters with closer pericenter distances, the star begins to shed mass and this mass loss further alters the orbital energy of the system. In the case where the star is only partially stripped (i.e., not destroyed completely as in Figure \ref{fig:full_disrupt}), the total orbital energy of the star after the first pericenter passage can be written as

\begin{equation}
\label{eq:Eorb}
E_{\rm{orb}} = E_{\rm{tides}} + E_{\rm{bind}} + E_{\rm{kick}}.
\end{equation}
$E_{\rm{tides}}$ is given by Equation~(\ref{eq:Etides}). $E_{\rm{bind}}$ is the binding energy of the disrupted star and ejected material:

\begin{equation}
    \label{eq:Ebind}
    E_{\rm{bind}} \approx - \frac{1}{2}\frac{G M_c \Delta M}{R_{\star}}
\end{equation}
where $M_c$ is the final mass of the stripped stellar core and $\Delta M$ is the total mass stripped from the star. $E_{\rm{kick}}$ is the kinetic energy imparted to the surviving stellar core from the mass loss. The origin of this ``kick'' comes from the third-order expansion of the gravitational potential which contains the asymmetries of the tidal field \citep[see, e.g.,][]{Brassart2008,Cheng2013}. As in \citet{Stone2013,Metzger2021}, we can estimate this term as

\begin{equation}
    \label{eq:Ekick}
    E_{\rm{kick}} \approx \frac{G M_{\rm{BH}} \Delta M}{r_p^3}R_{\star}^2.
\end{equation}
In the limit where $r_p < r_T$, $r_p$ is replaced with $r_T$ in this expression. As discussed in \citet{Metzger2021}, this simple expression agrees reasonably well with the energy computed from numerical simulations such as \citet{Manukian2013}.

If $E_{\rm{kick}} < \left| E_{\rm{tides}}+E_{\rm{bind}}\right|$, $E_{\rm{orb}}$ is negative and the partially disrupted remnant is bound to the black hole following the first encounter.\footnote{Of course, the process of tidal capture is also highly dependent on the relative velocity of the pair of objects. In the hyperbolic encounter regime, the general picture is expected to be quite different from the parabolic regime considered here.} Otherwise, the injection of energy into the star's orbit from the mass loss kick (after accounting for $E_{\rm{bind}}$) exceeds the removal of orbital energy through tides, and the net effect is an unbound star with, in principle, a high velocity kick. As can be read from the above equations, this outcome is qualitatively mostly likely to occur when $\Delta M$ is comparable to or larger than $M_c$.

\begin{figure}
    \centering
    \includegraphics[width=\columnwidth]{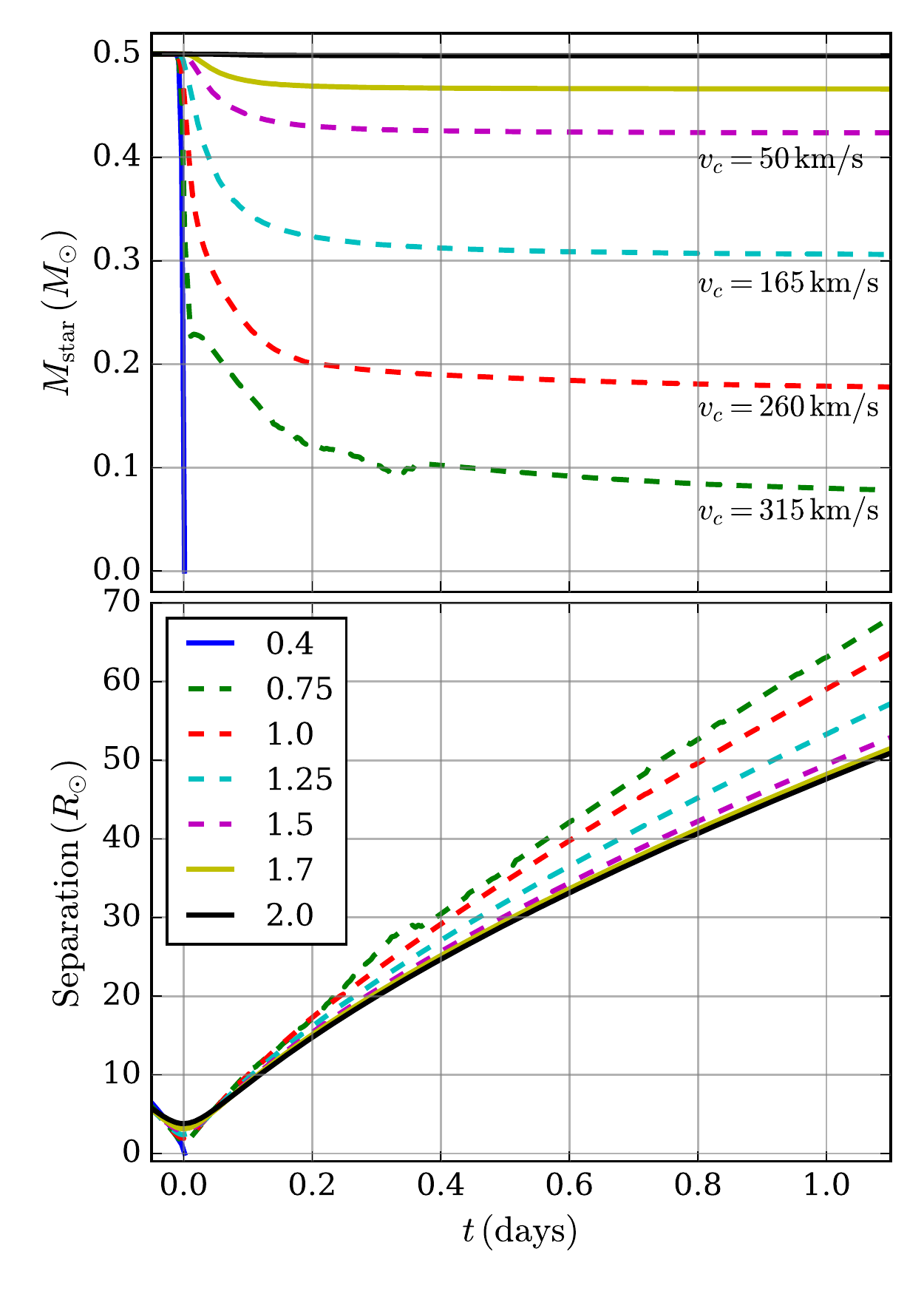}
    \caption{\footnotesize Stellar mass and orbital separation for encounters between $0.5\,M_{\odot}$ stars and $10\,M_{\odot}$ black holes for various pericenter distances (shown as different colors as indicated in the legend in bottom panel). For sufficiently small pericenter distance ($r_p/r_T \lesssim 0.4$), the star is disrupted fully on the first passage while for sufficiently wide pericenter distance ($r_p/r_T \gtrsim1.7$), little mass is stripped and the star is tidally captured by the black hole. For intermediate pericenter distances, the star receives an impulsive kick (with velocity values, $v_c$, indicated in the top panel) from asymmetric mass loss, as described in text.}
    \label{fig:evol_escapers}
\end{figure}

As shown in Figure~\ref{fig:summary}, the ``unbound'' outcome occurs uniquely in the simulations where $M_{\star} < 1\,M_{\odot}$ where the star is modeled as an $n=1.5$ polytrope. To determine whether the polytropic index (i.e., the stellar density profile) or the mass ratio of the interaction is the key factor that enables the ``unbound'' outcome, we ran two additional simulations with $r_p=r_T$. In the first, we adopt $M_{\rm{BH}}=10\,M_{\odot}$ and $M_{\star}=0.5\,M_{\odot}$ but with the star modeled as an Eddington standard model ($n=3$). In the second,  $M_{\rm{BH}}=20\,M_{\odot}$ and the star is a $M_{\star}=1\,M_{\odot}$ Eddington standard model. In both cases, the star is partially disrupted on the first passage but \textit{remains bound}. In this case, we conclude that, for the simulations considered here, the polytropic index is the determining factor for whether a partially disrupted star is bound or unbound after the first passage. This is reasonable: an $n=1.5$ polytrope has a more uniform mass distribution compared to an $n=3$ polytrope, which is more centrally concentrated. With a relatively large fraction of its mass found at larger radii, relatively more mass will be lost from $n=1.5$ polytrope at pericenter. In this case, $\Delta M$ is larger, and the impulsive kick received by the stellar remnant is larger. An $n=1.5$ polytrope is a reasonable model for sufficiently low-mass stars (e.g., low-mass M-dwarfs and brown dwarfs) which are expected to be deeply or fully convective \citep[e.g.,][]{HansenKawaler1994}. In this case, the unbound outcome is likely for realistic M-dwarf+black hole encounters.

The sole dependence of the boundedness of the core on the polytropic index is likely specific to our assumed mass ratios, which are relatively close to unity. Earlier work \citep[e.g.,][]{Manukian2013,Gafton2015} that explored mass ratios $q \ll 1$ demonstrated that unbound remnants can result for stars modeled with both $\Gamma = 4/3$ and $\Gamma = 5/3$ profiles. Furthermore, preliminary studies of the disruption of Eddington standard (and \texttt{MESA}) stellar models by intermediate-mass black holes ($M_{\rm BH} \gtrsim 100\,M_{\odot}$; $q \lesssim 10^{-2}$) using \texttt{StarSmasher} simulations similarly reveal unbound partially-disrupted cores for a range in $r_p$ values. In a forthcoming study (K{\i}ro\u{g}lu et al., in prep), we will explore how boundedness of partially disrupted cores changes as the mass ratio becomes increasingly far from unity.

In Figure \ref{fig:evol_escapers}, we show the time evolution of stellar mass and orbital separation for encounters between $0.5\,M_{\odot}$ stars and $10\,M_{\odot}$ black holes for various pericenter distances. For sufficiently small pericenter distance ($r_p/r_T \lesssim 0.4$), the star is disrupted fully on the first passage. For sufficiently wide pericenter distances ($r_p/r_T \gtrsim 1.7$), a small amount of mass is stripped from the star and the star is tidal captured. For intermediate pericenter distances ($r_p/r_T$ in the range $0.75-1.5$), sufficient mass is stripped and the partially disrupted star is kicked. In these cases, we indicate in the plot the final velocity (at infinity) of the stripped stellar core at the end of the SPH simulation.

As a robustness check, the final stellar velocities computed directly from the SPH models can be compared to velocity values estimated analytically from Equation~(\ref{eq:Eorb}). In the analytic case, the final velocity of the stripped stellar core is roughly $v_c \approx \sqrt{2 E_{\rm{orb}}/M_c}$, where $E_{\rm{orb}}$ is computed from Equations (\ref{eq:Etides}), (\ref{eq:Ebind}), and (\ref{eq:Ekick}). Inserting values for $r_p$,  $R_{\star}$, $M_{\star}$, $M_c$, and $\Delta M$ reported by our models ($\Delta M$ = $M_{\rm{BH,bound}}+M_{\rm{ej}}$; columns 5 and 7 in Table \ref{table:sims}), we can compute $v_c$ analytically. For example, for the case of $M_{\rm{BH}}=10\,M_{\odot}$, $M_{\star}=0.5\,M_{\odot}$, $r_p/r_T=1$ (simulation 15 in Table \ref{table:sims}) where $M_c \approx 0.2M_{\odot}$ and $\Delta M \approx 0.3M_{\odot}$, we compute $v_c\approx 230\,\rm{km/s}$, in reasonable agreement with the $260\,\rm{km/s}$ value reported in Figure \ref{fig:evol_escapers} that is read directly from the SPH result.

For a typical globular cluster of mass $M_{\rm{cl}}\sim10^6\,M_{\odot}$ and half-mass radius, $r_{\rm{h}}\sim 2\,$pc, the escape velocity is roughly $v_{\rm{esc}}\sim \sqrt{G M_{\rm{cl}}/r_{\rm{h}}}\sim 50\,\rm{km\,s}^{-1}$. Thus, for the cases shown as dashed curves in Figure \ref{fig:evol_escapers}, the kick imparted to the partially disrupted star is likely sufficient to eject the star from its host cluster.

\vspace{1cm}

\subsection{Momentum kick to center of mass of system}
\label{sec:COM_kick}

In addition to the momentum kick applied to the star from material stripped from the star itself, the center of mass of the entire black hole+star systems similarly receive a momentum kick as a result of asymmetric ejection of material from the system as a whole. As shown in column 7 of Table \ref{table:sims}, the amount of material unbound from the system can be significant in some cases. When the star is disrupted completely, this center-of-mass kick is applied to the black hole directly. In the partial disruption case, the kick is applied to the black hole+star pair (which may or may not be a bound binary).

The specific energy of the material ejected from the system can be approximated as

\begin{equation}
    \label{eq:Eej}
    \mathcal{E}_{\rm{ej}} \approx \frac{GM_{\rm{BH}}R_{\star}}{{r_T}^2} - \frac{1}{2}\frac{G(M_{\star}-M_{\rm{ej}})}{R_{\star}},
\end{equation}
where $M_{\rm{ej}}$ is the mass of material unbound from the system. For encounters with $r_p > r_T$, $r_T$ is replaced with $r_p$ in Equation~(\ref{eq:Eej}).
The characteristic velocity at infinity of the unbound material is then $v_{\rm{ej}} \approx \sqrt{2 \mathcal{E}_{\rm{ej}}}$ and the maximum linear momentum carried away by the unbound material is $|\textbf{P}|\sim M_{\rm{ej}}v_{\rm{ej}}$. In practice, the total momentum is less than this maximum because, in general, the ejected mass is not all moving in the same direction, however as a qualitative estimate, this is a reasonable approximation. In response, the center of mass of the remaining bound objects (either the black hole+partially disrupted star binary or isolated black hole+bound debris) receives a kick of velocity
\begin{equation}
    \label{eq:v_com}
    v_{\rm{com}} \approx \frac{M_{\rm{ej}}}{M_{\rm{BH}}+M_{\star}-M_{\rm{ej}}}v_{\rm{ej}}
\end{equation}
in the direction opposite the net momentum of the ejecta. Of course, for $M_{\rm{BH}}\gg M_{\rm{ej}}$, this kick is negligible, but when the ejected mass is comparable to the black hole mass (i.e., in the case of nearly head-on equal mass encounters; see Table \ref{table:sims}), the kick may be significant.

\begin{figure}
    \centering
    \includegraphics[width=\columnwidth]{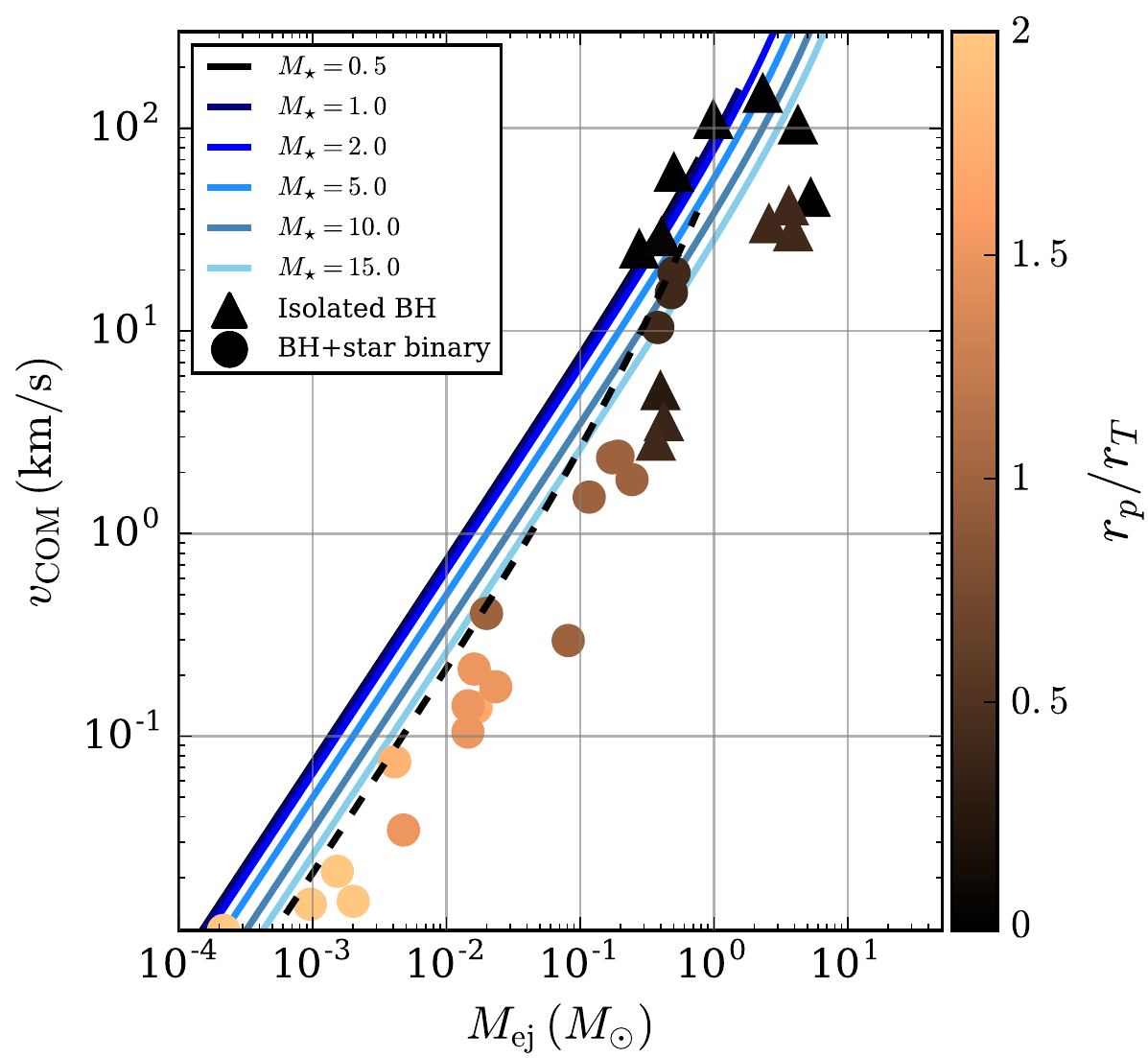}
    \caption{\footnotesize Velocity kick to the center of mass of final bound system, which depending on the encounter details can be either a black hole+partially disrupted star binary (circles) or an isolated black hole(+bound debris; triangles). Blue curves show the analytic scaling for the center-of-mass kick computed from Equations (\ref{eq:Eej}) and (\ref{eq:v_com}) for various stellar masses.}
    \label{fig:Vcom}
\end{figure}

In Figure \ref{fig:Vcom} we show the kick to the center of mass of the system versus total unbound ejecta mass after the first pericenter passage for all relevant SPH simulations (we exclude here simulations that lead to an unbound stellar core as described in Section \ref{sec:unbound_vs_bound} and simulations where a negligible amount of material is stripped from the star). In the case of partial disruption+tidal capture, the final system is a black hole+star binary (shown as circles in the figure). In the cases of a full disruption, the final system is an isolated black hole(+bound debris) (shown as triangles). Scatter points are colored according to the pericenter distance of each simulation. We show as colored blue curves the $v_{\rm{COM}}$ versus $M_{\rm{ej}}$ analytic relation computed using Equations (\ref{eq:Eej}) and (\ref{eq:v_com}) for various stellar masses $M_{\star}$ (noting that $R_{\star}$ and $r_T$ are both functions of $M_{\star}$ alone for a fixed $M_{\rm{BH}}=10\,M_{\odot}$ as assumed here). The dashed black curve shows the analytic relation for $M_{\star}=0.5\,M_{\odot}$ and $r_p=2r_T$ (the solid curves all adopt $r_p=r_T$), as an example of a more distant encounter. As noted previously, the analytic relation shown by the colored curves represents the \textit{maximum} COM velocity, assuming all material is ejected in same direction. This simple analytic relation reproduces well the results of the simulations to within a small factor.

As shown in the figure, more penetrating encounters lead to more mass loss and yield the highest kicks. In the most extreme case of a head-on collision (e.g., Figure \ref{fig:full_disrupt}), the black hole can receive a kick in excess of roughly 10'$\rm{s\,of\,km/s}$ or more, comparable to the maximum kick value estimated from Equation~(\ref{eq:v_com}) and potentially sufficient to eject the black hole from typical globular clusters.

Finally, we have discussed here center-of-mass velocities following only the first pericenter passage. As shown in Figure \ref{fig:partial_bound}, it is likely that with multiple passages, the final outcome of these encounters is for all the stellar material (including the tidally captured star) bound to the black hole to be accreted or ejected. If any subsequent mass loss during later pericenter passages carries no net linear momentum, then the $v_{\rm{COM}}$ values reported in Figure~\ref{fig:Vcom} can be read as the final black hole velocity with respect to the initial center of mass before the encounter.

\bigskip

\subsection{Multiple passages}
\label{sec:multiple_passage}

As shown in Figures \ref{fig:partial_bound} and \ref{fig:evol_insp}, if a partially disrupted stellar core is tidally captured by the black hole during the first pericenter passage, the core will return to pericenter, undergoing additional disruption(s). This process continues until the core inspirals completely and is fully disrupted by the black hole. The process of partial disruption and tidal capture of stars by stellar black holes has been explored previously for disruptions of main-sequence and giant-branch stars \citep[e.g.,][]{Godet2014,Ivanova2017} and in other contexts \citep[e.g.,][]{Rosswog2004}.

Modeling the full inspiral process through all pericenter passages is computationally challenging, particularly in cases where the time for the stellar core to return to pericenter is very long (e.g., months or more; see Figure \ref{fig:summary}). In this study, we focus primarily upon the outcome of the first passage and therefore stop most of our simulations after the first passage, even if the stellar core is bound and would return to pericenter in reality. However, to get a sense of the ultimate outcome in the case of multiple passages, we run a select few models all the way to inspiral.

\begin{figure*}
    \centering
    \includegraphics[width=0.9\linewidth]{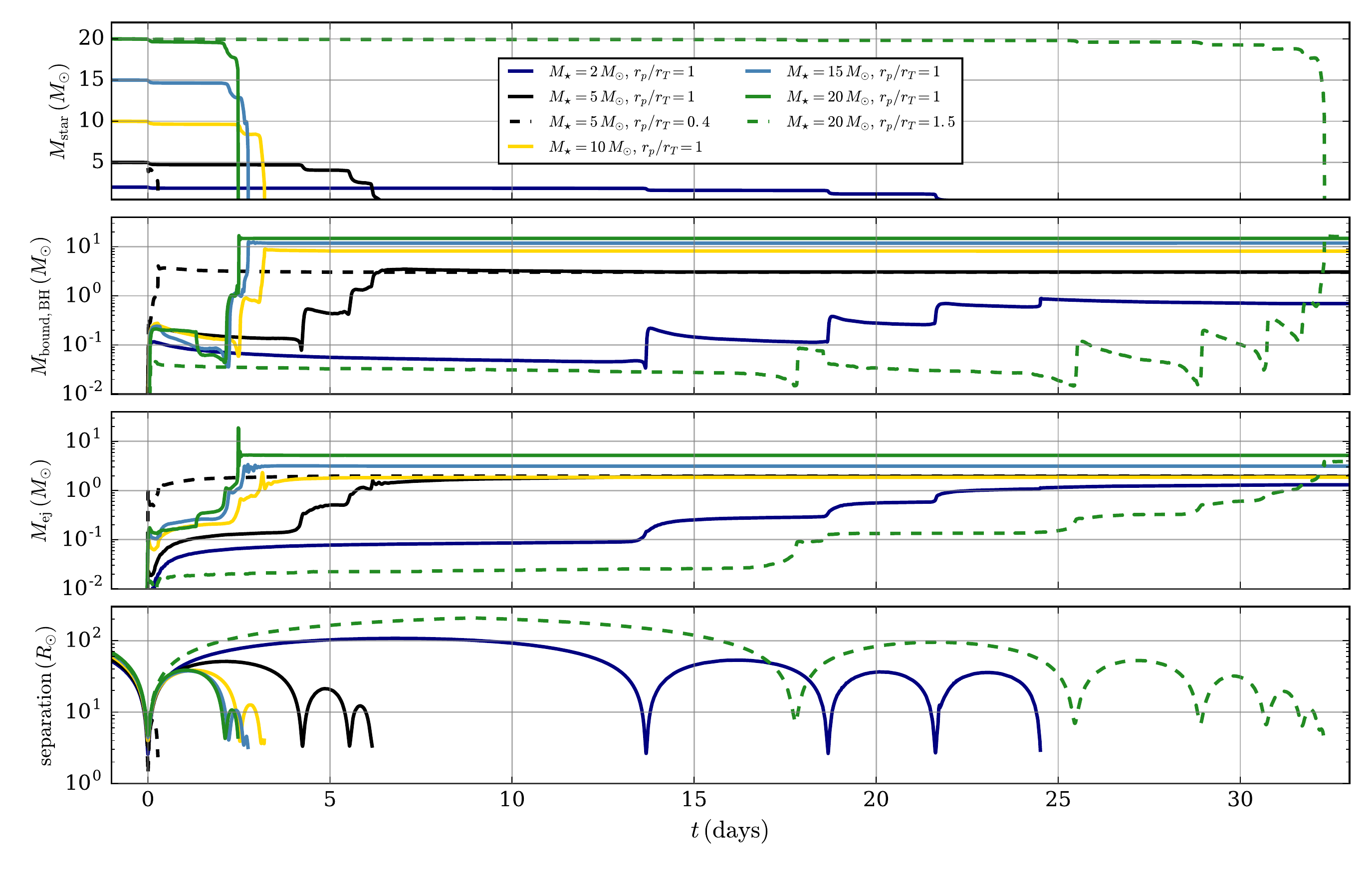}
    \caption{\footnotesize Analogous to Figure \ref{fig:evol_insp} but for a wider range of systems. Different colored curves show simulations with different initial stellar masses (in all cases we assume a black hole mass of $10\,M_{\odot}$). The solid curves show simulations with $r_p/r_T=1$, the dashed black curve shows an example of a more penetrating interaction ($r_p/r_T=0.4$), and the dashed green curve shows an example of a less penetrating interaction ($r_p/r_T=1.5$). In general, higher mass ratios and higher penetration factors lead to more mass loss at first pericenter passages, shorter timescale for the partially disrupted remnant to return to pericenter a second time, and fewer passages overall before the star is disrupted fully. }
    \label{fig:evol_insp_all}
\end{figure*}

The specific number of passages and time between passages varies with the mass ratio and pericenter distance. We demonstrate this in Figure \ref{fig:evol_insp_all}. Here we show the stellar mass, mass bound to black hole, ejecta mass, and separation versus time for several models. In dark blue, black, yellow, light blue, and green we show simulations with initial stellar mass of $2\,M_{\odot}$, $5\,M_{\odot}$, $10\,M_{\odot}$, $15\,M_{\odot}$, and $20\,M_{\odot}$, respectively all for initial pericenter passages of $r_p=r_T$. The $5\,M_{\odot}$ case (black curve) is identical to the simulation shown in Figures \ref{fig:partial_bound} and \ref{fig:evol_insp}. As shown, for a fixed $r_p/r_T$, higher mass ratios ($M_\star/M_{\rm{BH}}$) lead to shorter times for the partially disrupted star to return to pericenter and fewer passages until full disruption.

Dashed curves in Figure \ref{fig:evol_insp_all} demonstrate the effect of the pericenter distance. As dashed black curves, we show the evolution of a $5\,M_{\odot}$ star with $r_p/r_T=0.4$. After the first passage, roughly $1.2\,M_{\odot}$ of material is stripped from the star and the disrupted star returns to pericenter after roughly 6 hours after which the star is disrupted fully. This is in contrast to the $r_p/r_T=1$ case (solid black curve), where only $0.3\,M_{\odot}$ of material is stripped from the star and the disrupted star returns to pericenter after roughly $4\,$days. Meanwhile, dashed green curves show an example of a less penetrative interaction. In dashed green, we show the evolution of a $20\,M_{\odot}$ star with $r_p/r_T=1.5$. After the first passage, roughly $0.05\,M_{\odot}$ is stripped from the star (compared to roughly $0.3\,M_{\odot}$ in the $r_p/r_T=1$ case; solid green curve) and the time to return for second pericenter passage is roughly $8\,$days (compared to $2\,$days in $r_p/r_T=1$ case). In general, we find that for a fixed mass ratio, more penetrative interactions lead to more mass loss at pericenter, shorter timescales for the return to pericenter, and fewer passages before the star is disrupted fully.

\startlongtable
\begin{deluxetable*}{l|cc|cc|ccc|ccccc|c}
\tabletypesize{\footnotesize}
\tablewidth{0pt}
\tablecaption{List of simulations performed \label{table:sims}}
\tablehead{
	\colhead{} &
	\colhead{$^1M_{\rm{BH}}$} &
	\colhead{$^2M_{\star,i}$} &
	\colhead{$^3r_p/R_{\star}$} &
	\colhead{$^4r_p/r_T$} &
	\colhead{$^5M_{\rm{bound,BH}}$} &
	\colhead{$^6M_{\star,f}$} &
	\colhead{$^7M_{\rm{ej}}$} &
	\colhead{$^8t_{\rm{orb}}$} &
	\colhead{$^9J_{\rm{disk}}$} &
	\colhead{$^{10}R_{\rm{disk}}$} &
	\colhead{$^{11}\Omega_{\rm{disk}}$} &
	\colhead{$^{12}t_{\rm{acc}}$} &
	\colhead{$^{13}P_{\rm{orb}}$}\\
	\colhead{} &
	\multicolumn{2}{c}{$M_{\odot}$} &
	\colhead{} &
	\colhead{} &
	\multicolumn{3}{c}{$M_{\odot}$} &
	\colhead{hours} &
	\colhead{$M_{\odot} R_{\odot}^2 \rm{d}^{-1}$} &
	\colhead{$R_{\odot}$} &
	\colhead{$\rm{d}^{-1}$} &
	\colhead{days} &
	\colhead{days}
}
\startdata
1& 30 & 0.5 & 0.10 & 0.03 & 0.119    & 0.000   & 0.381    & 2.30    & 15.5      & 7.90    & 2.09    & 4.79     & N/A      \\ 
2& 30 & 0.5 & 0.70 & 0.18 & 0.245    & 0.000   & 0.255    & 2.30    & 67.5      & 16.8    & 0.97    & 10.3     & N/A      \\ 
3& 30 & 0.5 & 1.10 & 0.28 & 0.270    & 0.000   & 0.230    & 2.30    & 82.0      & 19.0    & 0.84    & 11.9     & N/A      \\ 
4& 30 & 0.5 & 1.59 & 0.41 & 0.286    & 0.000   & 0.214    & 2.30    & 100       & 18.8    & 1.00    & 10.0     & N/A      \\ 
5& 30 & 0.5 & 2.71 & 0.69 & 0.272    & 0.000   & 0.228    & 2.30    & 124       & 17.2    & 1.55    & 6.45     & N/A      \\ 
6& 30 & 0.5 & 3.91 & 1.00 & 0.246    & 0.152   & 0.102    & 2.30    & 129       & 16.2    & 2.00    & 4.99     & N/A      \\ 
7& 30 & 0.5 & 5.87 & 1.50 & 0.059    & 0.425   & 0.016    & 4.23    & 31.6      & 14.9    & 2.42    & 4.13     & N/A      \\ 
8& 30 & 0.5 & 7.83 & 2.00 & $<$0.001 & 0.50    & $<$5e-5  & 6.52    & -         & -       & -       & -        & 800      \\ 
\hline
 9& 10 & 0.5 & 0.00 & 0.00 & 0.224    & 0.000   & 0.276    & 2.30    & -         & -       & -       & -        & N/A      \\ 
10& 10 & 0.5 & 0.10 & 0.04 & 0.177    & 0.000   & 0.323    & 2.30    & 11.8      & 5.31    & 2.36    & 4.24     & N/A      \\ 
11& 10 & 0.5 & 0.70 & 0.26 & 0.299    & 0.000   & 0.201    & 2.30    & 45.3      & 12.0    & 1.04    & 9.57     & N/A      \\ 
12& 10 & 0.5 & 1.00 & 0.37 & 0.322    & 0.000   & 0.178    & 2.30    & 50.8      & 12.6    & 0.99    & 10.1     & N/A      \\ 
13& 10 & 0.5 & 1.10 & 0.41 & 0.322    & 0.000   & 0.178    & 2.30    & 51.9      & 12.5    & 1.03    & 9.68     & N/A      \\ 
14& 10 & 0.5 & 2.04 & 0.75 & 0.257    & 0.123   & 0.120    & 2.30    & 50.7      & 9.89    & 2.01    & 4.96     & N/A      \\ 
15& 10 & 0.5 & 2.71 & 1.00 & 0.232    & 0.202   & 0.066    & 2.30    & 49.7      & 9.24    & 2.51    & 3.98     & N/A      \\ 
16& 10 & 0.5 & 3.39 & 1.25 & 0.143    & 0.318   & 0.039    & 3.22    & 30.3      & 8.17    & 3.17    & 3.15     & N/A      \\ 
17& 10 & 0.5 & 4.07 & 1.50 & 0.064    & 0.426   & 0.010    & 4.23    & 14.1      & 7.19    & 4.25    & 2.35     & N/A      \\ 
18& 10 & 0.5 & 4.48 & 1.65 & 0.031    & 0.467   & 0.002    & 4.88    & 7.05      & 6.97    & 4.67    & 2.14     & 303      \\ 
19& 10 & 0.5 & 4.75 & 1.75 & 0.017    & 0.483   & 0.0004   & 5.33    & 3.88      & 7.13    & 4.54    & 2.20     & 171      \\ 
20& 10 & 0.5 & 5.43 & 2.00 & $<$0.002 & 0.50    & $<$1e-4  & 6.52    & -         & -       & -       & -        & 265      \\ 
\hline
21& 10 & 1   & 0.00 & 0.00 & 0.499    & 0.000   & 0.501    & 2.78    & -         & -       & -       & -        & N/A      \\ 
22& 10 & 1   & 0.88 & 0.41 & 0.435    & 0.300   & 0.266    & 2.78    & 52.0      & 6.33    & 2.99    & 3.34     & 3.29     \\ 
23& 10 & 1   & 2.15 & 1.00 & 0.056    & 0.932   & 0.012    & 2.78    & 10.7      & 3.93    & 12.3    & 0.82     & 40.0     \\ 
24& 10 & 1   & 3.23 & 1.50 & 0.002    & 0.998   & 0.0002   & 5.11    & 0.489     & 6.83    & 4.58    & 2.18     & 1200     \\ 
25& 10 & 1   & 4.31 & 2.00 & $<$1e-5  & 1.000   & $<$5e-6  & 7.87    & -         & -       & -       & -        & $7.47\times 10^4$ \\ 
\hline
26& 10 & 2   & 0.00 & 0.00 & 1.012    & 0.000   & 0.988    & 3.61    & -         & -       & -       & -        & N/A      \\ 
27& 10 & 2   & 0.70 & 0.41 & 0.696    & 0.823   & 0.481    & 3.61    & 103       & 9.23    & 1.73    & 5.78     & 0.636    \\ 
28& 10 & 2   & 1.71 & 1.00 & 0.110    & 1.870   & 0.020    & 3.61    & 21.4      & 4.80    & 8.47    & 1.18     & 13.9     \\ 
29& 10 & 2   & 2.57 & 1.50 & 0.006    & 1.994   & 0.0006   & 6.64    & 1.20      & 6.77    & 4.58    & 2.19     & 515      \\ 
30& 10 & 2   & 3.42 & 2.00 & $<$2e-5  & 2.000   & $<$5e-6  & 10.2    & -         & -       & -       & -        & $3.49\times 10^4$ \\ 
\hline
31& 10 & 5   & 0.00 & 0.00 & 2.68     & 0.000   & 2.32     & 5.21    & -         & -       & -       & -        & N/A      \\ 
32& 10 & 5   & 0.52 & 0.41 & 3.70     & 0.000   & 1.30     & 5.21    & 905       & 9.60    & 2.65    & 3.77     & N/A      \\ 
33& 10 & 5   & 1.26 & 1.00 & 0.214    & 4.728   & 0.058    & 5.21    & 43.9      & 7.30    & 3.85    & 2.60     & 4.27     \\ 
34& 10 & 5   & 1.89 & 1.50 & 0.017    & 4.980   & 0.003    & 9.58    & 3.83      & 10.1    & 2.16    & 4.64     & 142      \\ 
35& 10 & 5   & 2.52 & 2.00 & $<$5e-05 & 5.000   & $<$5e-4  & 14.7    & -         & -       & -       & -        & $2.49\times 10^4$ \\ 
\hline
36& 10 & 10  & 0.00 & 0.00 & 5.76     & 0.000   & 4.24     & 7.04    & -         & -       & -       & -        & N/A      \\ 
37& 10 & 10  & 0.41 & 0.41 & 7.65     & 0.000   & 2.35     & 7.04    & 1700      & 10.4    & 2.06    & 4.86     & N/A      \\ 
38& 10 & 10  & 1.00 & 1.00 & 0.210    & 9.643   & 0.147    & 7.04    & 43.9      & 10.8    & 1.80    & 5.56     & 2.54     \\ 
39& 10 & 10  & 1.50 & 1.50 & 0.033    & 9.961   & 0.006    & 12.9    & 7.55      & 14.1    & 1.14    & 8.78     & 55.7     \\ 
40& 10 & 10  & 2.00 & 2.00 & $<$1e-4  & 10.00   & $<$0.001 & 19.9    & -         & -       & -       & -        & 3200     \\ 
\hline
41& 10 & 15  & 0.00 & 0.00 & 9.71     &  0.000 & 5.29     & 8.27    & -         & -       & -       & -        & N/A      \\ 
42& 10 & 15  & 0.36 & 0.41 & 11.34    &  0.000 & 3.66     & 8.27    & 2160      & 11.0    & 1.56    & 6.42     & N/A      \\ 
43& 10 & 15  & 0.87 & 1.00 & 0.137    & 14.646 & 0.217    & 8.27    & 29.0      & 13.4    & 1.17    & 8.51     & 2.23     \\ 
44& 10 & 15  & 1.31 & 1.50 & 0.036    & 14.950 & 0.014    & 15.2    & 6.83      & 18.2    & 0.57    & 17.5     & 35.0     \\ 
45& 10 & 15  & 1.75 & 2.00 & $<$2e-4  & 15.00  & $<$0.002 & 23.4    & -         & -       & -       & -        & 1370      \\ 
\hline
46& 10 & 20  & 0.32 & 0.40 & 16.05    & 0.0000 & 3.95     & 9.14    & 2790      & 11.5    & 1.31    & 7.64     & N/A      \\ 
47& 10 & 20  & 0.79 & 1.00 & 0.202    & 19.633 & 0.165    & 9.14    & 34.0      & 11.5    & 1.28    & 7.82     & 2.12     \\ 
48& 10 & 20  & 1.19 & 1.50 & 0.037    & 19.944 & 0.019    & 16.8    & 7.13      & 19.1    & 0.53    & 18.8     & 26.0     \\ 
49& 10 & 20  & 1.59 & 2.00 & $<$2e-4  & 20.00  & $<$0.005 & 25.9    & -         & -       & -       & -        & 716      \\ %
\enddata
\tablecomments{\footnotesize List of all simulations performed in this study. In columns 1-4, we list initial conditions for the simulations. In columns 5-7, we list the total mass bound to the black hole, the final stellar mass, and the total mass unbound from the system after the first pericenter passage. In column 8, we list the orbital time characterizing the timescale for bound material to return to pericenter (see Equation~(\ref{eq:t_fb}). In columns 9-11 we list various properties of the disk of material surrounding the black hole (when applicable). These disk properties (as well as the properties in columns 5-7) are all reported at $2t_{\rm{orb}}$ after the first pericenter passage, as described in the text. In column 12 we list the viscous accretion time scale for the disk to accrete onto the black hole. Finally, in column 13 we list the orbital period of the partially disrupted star to return to pericenter (in cases where the star is not fully disrupted and where the star becomes bound to the black hole).}
\end{deluxetable*}

\subsection{Disk properties}

In all cases where a fraction of material is torn from the star and becomes bound to the black hole, the characteristic timescale for bound material to return to pericenter can be estimated simply as the orbital period of the stripped material bound to the black hole

\begin{figure}
    \centering
    \includegraphics[width=\linewidth]{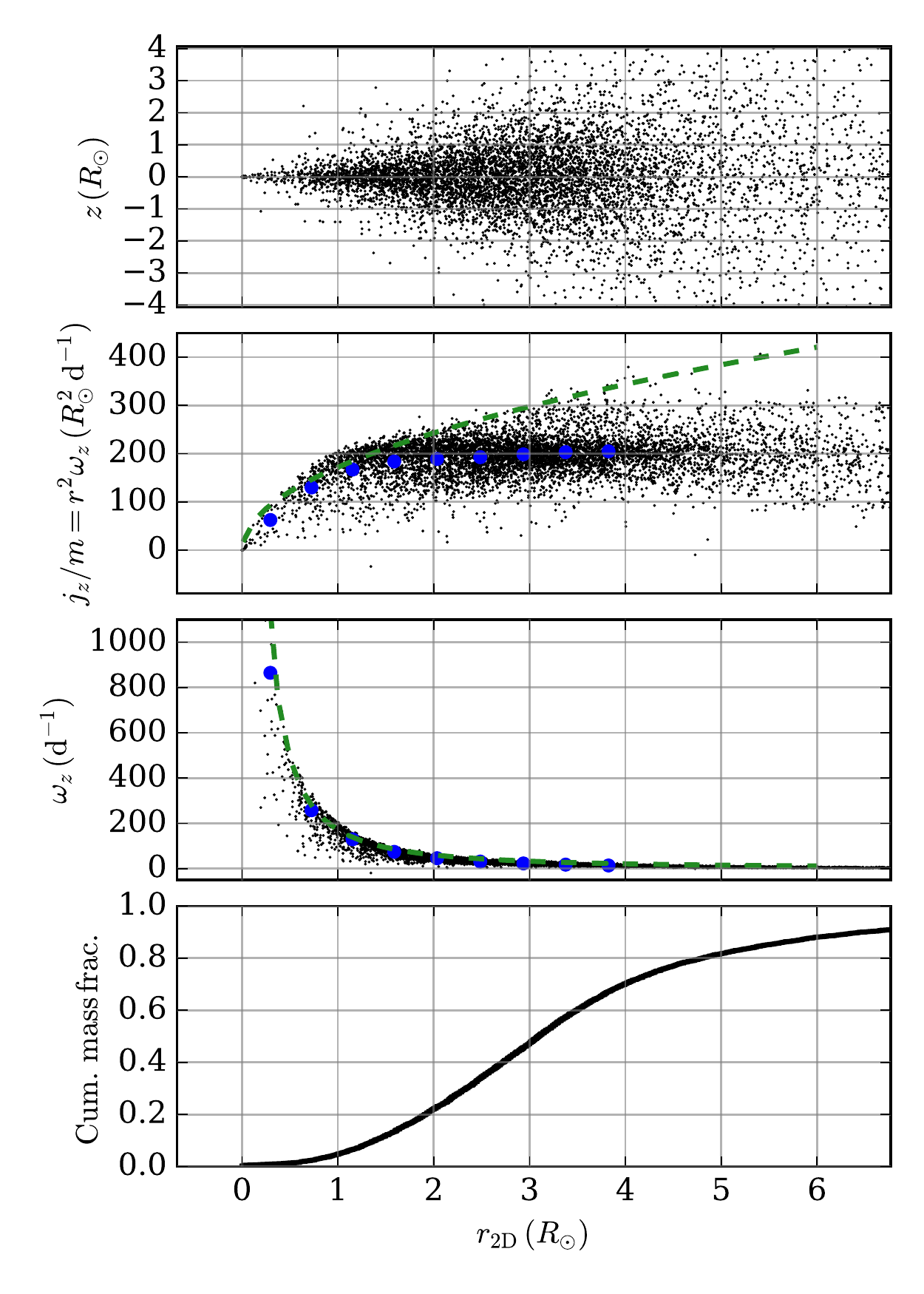}
    \caption{\footnotesize All SPH particles (black scatter points) bound to the black hole in the snapshot roughly hours ($2t_{\rm{orb}}$; see Equation \ref{eq:t_fb}) after the first pericenter passage for the $2\,M_{\odot}$ star+$10\,M_{\odot}$ black hole simulation with $r_p/r_T=1$ (chosen as a representative example of a partial tidal disruption). Top panel shows the $z$ position versus 2D radial position (with $r_{\rm{2D}}=0$ defined at the black hole location) for all SPH particles bound to the black hole. Second panel shows the specific angular momentum of all particles. Third panel shows the angular velocity about $z$-axis. The dashed green curves in the second and third panels show for reference the Keplerian profiles. The bottom panel shows the cumulative mass fraction of all material bound to the black hole. For reference, blue circles show the average values within 9 radial bins from $r=0-2r_p$.}
    \label{fig:disk}
\end{figure}

\begin{equation}
    \label{eq:t_fb}
    t_{\rm orb} = 2\pi \sqrt{\frac{\max(r_p^3,r_T^3)}{G M_{\rm BH}}} \approx 3 \Bigg( \frac{M_{\rm BH}}{10\,M_{\odot}} \Bigg)^{-1/2} \Bigg( \frac{r_p}{2.2\,R_{\odot}} \Bigg)^{3/2} \, \rm{hr} 
\end{equation}
where the fiducial value of $r_p=2.2R_{\odot}$ is the tidal disruption radius of a $1M_{\odot}$ star for a $10M_{\odot}$ black hole.
For $r_p > r_T$, the orbital timescale
increases; however, for $r_p = r_T$, a minimum value is
achieved since more penetrating encounters do not yield
shorter times for the stripped debris to orbit the black hole.
Note this timescale differs from the characteristic fallback timescale $t\propto r_p^3$ often used in the supermassive black hole TDE literature computed under the analytic ``frozen in'' approximation \citep[e.g.,][]{Lacy1982,Rees_1988,EvansKochanek1989}. This approximation, appropriate for $R_{\star}\ll r_p$ as in the supermassive black hole TDE regime, assumes the star is a static sphere that is instantaneously destroyed upon reaching the tidal radius. However, in the limit considered here where $R_{\star} \sim r_p$, this approximation is not necessarily valid and we find the $t \propto r_p^{3/2}$ scaling in Equation~(\ref{eq:t_fb}) more effectively captures the timescale on which the debris stream encircles the black hole. Values of $t_{\rm orb}$ for our simulations are listed in column 8 of Table~\ref{table:sims}.

On a timescale comparable to $t_{\rm{orb}}$, the debris stream self-intersects. As a result, a fraction of the bound debris becomes unbound through energy acquired through shocks, while the remaining material circularizes and forms a thick rotating torus of radius $r_d\sim 2r_p$. Throughout this paper, we refer to these thick torii of bound material as a ``disks'' for simplicity. However, the accretion physics of these rotating envelopes of material bound to the black hole are uncertain and may be quite different from that of a classical ``thin'' accretion disk \citep[e.g.,][]{Frank2002}. We can, nonetheless, use some of the standard accretion disk nomenclature to inform the basic properties of these disks and their possible electromagnetic signatures.

Once a disk is formed, the material accretes onto the black hole on a viscous accretion timescale. This general picture holds for all pericenter distances with the exception of perfectly head-on collisions with $r_p=0$ where the total angular momentum of the black hole+star system is zero. In this limiting case, a disk does not form and the bound material simply accretes on a freefall time.

In this study, we consider gravity and hydrodynamics alone and do not consider accretion processes. However, by analyzing the structure of the material bound to the black holes (e.g., the disk) at the end of our hydrodynamic simulations, we can attain a rough sense of the accretion physics that may be at play. In Figure \ref{fig:disk}, we show a representative example of the properties of the material bound to the black hole after the disruption. Here we show all SPH particles after $2t_{\rm{orb}}$ (to ensure a disk has had sufficient time to form) for the case of a $2\,M_{\odot}$ star interacting with a $10\,M_{\odot}$ black hole at $r_p=r_T$. This particular model is chosen simply as a representative case. All partial disruptions exhibit qualitatively similar behavior. In the top panel we show the $z$ position (for a disruption occurring in the $xy$ plane) versus 2D radial position (with $r_{\rm{2D}}=0$ defined at the black hole location) for all SPH particles bound to the black hole. This gives a visual representation of the disk geometry. The second panel shows the specific angular momentum ($j_z/m$, where $m$ is the particle mass) of all SPH particles. The third panel shows the angular velocity about the $z$-axis for all particles. The dashed green curves in the second and third panels show for reference the Keplerian profiles ($j\propto r^{1/2}$ and $\omega \propto r^{-3/2}$, respectively). As shown the disk profile is nearly Keplerian at small radii but begins to diverge from Keplerian at larger radii. 
Finally, in the bottom panel we show the cumulative mass fraction of all material bound to the black hole.

From the SPH snapshots as in Figure \ref{fig:disk}, we can calculate characteristic disk properties that can then be used to estimate the characteristic viscous accretion time scale for material bound to the black hole after the first pericenter passage. The total angular momentum $J_z$ around the z-axis (that is, the axis perpendicular to the orbital plane passing through the black hole) is computed trivially by adding the angular momenta of all SPH particles bound to the black hole

\begin{equation}
    J_z = \sum_i m_i r_i^2 \omega_z^i,
\end{equation}
where $\omega_z^i$ is the angular velocity about the z-axis for particle $i$ (see, e.g., third panel of Figure \ref{fig:disk}). The characteristic disk radius is computed as a mass-weighted average of all bound particles:

\begin{equation}
    R_{\rm{disk}} = \sqrt{I_z/M_{\rm{disk}}}
\end{equation}
where $I_z=\sum m_i r_i^2$ is the moment of inertia about the z-axis of the disk and $M_{\rm{disk}}$ is the total disk mass. Finally, the characteristic angular velocity of the disk is computed as

\begin{equation}
    \Omega_{\rm{disk}} = J_z/I_z.
\end{equation}
In the standard \citet{ShakuraSunyaev1973} model, the characteristic viscous accretion timescale can be estimated as
\begin{equation}
    \label{eq:t_acc}
    t_{\rm{acc}} = [\alpha h^2 \Omega]^{-1}
\end{equation}
where $\alpha$ is the dimensionless viscosity parameter (here we assume $\alpha=0.1$) and $h = H/R_{\rm{disk}}$ (where $H$ is the disk scale height). Here, we simply assume $h\approx 1$, motivated by results of the models (e.g., see top panel of Figure \ref{fig:disk}).

In columns 9-12 of Table \ref{table:sims}, we show disk properties after the first pericenter passage for all simulations. Disk properties are reported at $2t_{\rm{orb}}$ after the first pericenter passage, with $t_{\rm{orb}}$ given by Equation~(\ref{eq:t_fb}). This time is chosen to ensure the disk has had sufficient time to form. On longer timescales, viscous accretion processes are expected to operate that are not modeled in our current SPH models. In Section \ref{sec:EM}, we extrapolate basic features expected from the accretion process on longer timescales based on the properties of the disks at formation that are reported here.

Of course, for perfectly head-on collisions, the disk properties are undefined since by definition the angular momentum about z-axis is zero. In general, we find viscous accretion timescales of roughly $1-10\,$days, roughly an order of magnitude larger than the typical fallback timescales which range from roughly $0.1-1\,$days (see column 8 of Table \ref{table:sims}). Thus, as predicted by analytic results of \citet{Perets_2016,Kremer2019c}, the viscous accretion timescale is likely the key timescale for determining the evolution of the associated EM transient. This is in contrast to classical supermassive black hole TDEs which are dominated by the fallback timescale \citep[e.g.,][]{Rees_1988}. 

\section{Electromagnetic signatures}
\label{sec:EM}

Although detailed treatment of radiation, accretion, and accretion feedback is beyond the scope of this study which focuses on the hydrodynamics of the problem, we can examine in post-processing possible electromagnetic signatures that may result from these interactions using analytic arguments informed from the results of our SPH models. In particular, the properties of the material bound to the black hole (total mass, angular momentum, etc.) at the end of the SPH simulations can help inform the possible outcomes. In the following subsections, we discuss the various electromagnetic signatures that may result from the various regimes explored.

\subsection{Transient signatures from first passage}
\label{sec:first_passage}

Two characteristic timescales are useful for determining the typical timescales for accretion of bound material onto the black hole: the fallback time (typically hours; see Equation \ref{eq:t_fb}) and the viscous accretion timescale (typically days; see Equation \ref{eq:t_acc}). In cases where the black hole and star collide nearly head on ($r_p\sim0$), the angular momentum of the debris that becomes bound to the black hole is negligible and the fallback time is more relevant. In the more general case of off-center interactions, the debris forms a rotating torus (``disk'') around the black hole and the viscous accretion timescale is relevant. The characteristic mass is the total material bound to the black hole at the end of a given pericenter passage, labelled $M_{\rm{disk}}$. Depending on the initial stellar mass and the distance of closest approach, the total mass bound to the black hole after the first passage ranges from roughly $10^{-3}-\rm{a\, few}\,\it{M}_{\odot}$ (see column 5 of Table \ref{table:sims}).

The characteristic accretion rate onto the black hole can then be written as $\dot{M}\sim M_{\rm disk}/\Delta t$, where $\Delta t$ is the characteristic accretion timescale. For off-center interactions where the total angular momentum of material bound to the black hole is nonzero, $\Delta t$ is the viscous accretion time of the disk, $t_{\rm{acc}}$ (Equation~\ref{eq:t_acc}; column 12 of Table~\ref{table:sims}). For the specific case of a head-on collision with $r_p=0$, material can fall directly onto the black hole and $\Delta t$ is more appropriately given by the freefall time. For $M_{\rm disk} \approx 10^{-3}-10\,M_{\odot}$ and $\Delta t\approx1-10\,$d, $\dot{M}$ ranges from roughly $10^{-2}-10^3\,M_{\odot}\,\rm{yr}^{-1}$. Such values are several orders of magnitude in excess of the classic Eddington accretion limit, $\dot{M}_{\rm Edd} \approx 2\times10^{-7} (M_{\rm BH}/10\,M_{\odot}) M_{\odot}\,\rm{yr}^{-1}$ adopting a radiative efficiency of 0.1 and taking $L_{\rm Edd} \approx 10^{39} (M_{\rm BH}/10\,M_{\odot}) \, \rm{erg\,s}^{-1}$ as the electron-scattering Eddington luminosity. For such high mass inflow rates, photons are trapped and advected inward \citep[e.g.,][]{Begelman1979}. In this ``hypercritical'' accretion regime, since the disk is unable to cool efficiently via radiation, it is susceptible to outflows that ultimately reduce the total mass supplied to the black hole \citep[e.g.,][]{NarayanYi1995,Blandford1999}.

To account for the uncertainties pertaining to such outflows, we assume \citep[as in our previous studies; e.g.,][]{Kremer2019c} that the accretion rate onto the black hole is reduced by a factor $(R_{\rm in}/R_d)^s$, where $R_{\rm in}=6GM_{\rm BH}/c^2$ is the inner edge of the disk (assumed to be the innermost stable circle orbit), $R_d$ is the outer edge of the disk, and the power-law index $s \in [0,1]$. This treatment is in line with that of previous studies of thick super-Eddington accretion disks \citep[e.g.,][]{Blandford1999}.
In the case of highest mass inflow rate, $s=0$, the entire disk mass is accreted onto the black hole on roughly a viscous time, yielding a peak accretion rate $\dot{M}\sim M_{\rm disk}/t_{\rm{acc}}$. In the case of lowest mass inflow rate, $s=1$, the accretion rate is reduced by a factor of roughly $10^{-5}$. Numerical simulations of radiatively inefficient accretion flows show $s$ most likely lies in the range $0.2-0.8$ \citep[e.g.,][]{Yuan2012,Yuan2014}.

The peak accretion luminosity from the inner disk near the accretion radius can be estimated as

\begin{equation}
    \label{eq:Lum}
    L_{\rm{peak}} \approx \epsilon \dot{M}_{\rm{BH}} c^2 \approx \epsilon \frac{M_{\rm{disk}}}{\Delta t}\Bigg(\frac{R_{\rm in}}{R_d} \Bigg)^s c^2
\end{equation}
where $\epsilon$ is the accretion efficiency factor. Previous studies of super-Eddington accretion flows onto black holes \citep[e.g.,][]{Ohsuga2009,Jiang2014,CoughlinBegelman2014,SadowskiNarayan2015,Takahashi2016,CoughlinBegelman2020,Kitaki2021} suggest relativistic outflows collimated along the disk rotation axis (e.g., a jet-like geometry) can be launched from the inner disk, potentially enabling a large fraction of the accretion energy to be released. Radiation GRMHD simulations \citep[e.g.,][]{SadowskiNarayan2016} have confirmed this basic structure and identified accretion efficiency factors of $\epsilon \approx 10^{-2}$ for accretion rates a few hundred times higher than the Eddington rate. We adopt $\epsilon \approx 10^{-2}$ as our fiducial efficiency, but note that the accretion rates modeled in the aforementioned studies are still several orders of magnitude lower than the accretion rates predicted here. In this case, the accretion luminosities quoted below may be viewed as upper limits.

\begin{figure}
    \centering
    \includegraphics[width=\columnwidth]{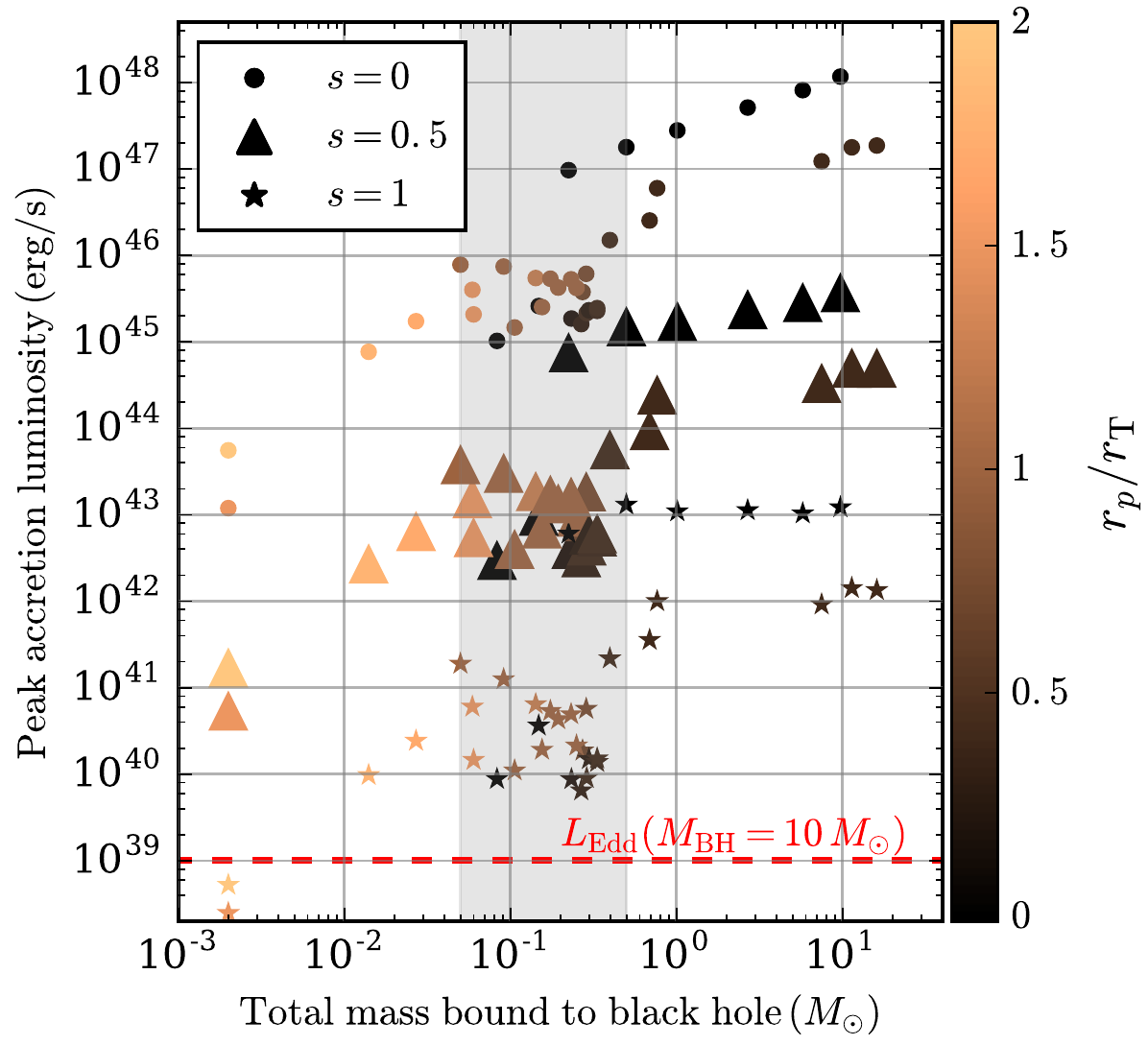}
    \caption{\footnotesize Peak luminosity from accretion onto black hole using Equation~(\ref{eq:Lum}) versus total mass of disrupted material bound to the black hole after the first passage. As in Figure \ref{fig:Vcom}, color denotes the penetration factor of the interaction. As different symbols, we show the peak luminosity calculated for different values of the $s$ exponent in Equation~(\ref{eq:Lum}) which parameterizes the fraction of material transported from the outer edge of the disk to the black hole. $s=0$ (circles) is the most optimistic case, $s=1$ (stars) is most pessimistic, and $s=0.5$ (large triangles) is most realistic \citep[see, e.g.,][]{Yuan2012}. The gray background marks the most typical disk mass for the most common types of interaction (discussed further in Section \ref{sec:rates}. For reference, the dashed red line indicates the Eddington luminosity for a $10\,M_{\odot}$ black hole.}
    \label{fig:lum}
\end{figure}

In Figure \ref{fig:lum}, we show the peak accretion luminosity versus total mass bound to the black hole after the first pericenter passage for all simulations in this study. We use Equation~(\ref{eq:Lum}): $M_{\rm{disk}}$ (column 5 of Table \ref{table:sims}), $r_d$ (column 10 of Table \ref{table:sims}), and $\Delta t$ are obtained directly from the simulations. For $\Delta t$, we use the viscous accretion timescale, $t_{\rm{acc}}$ computed for each simulation (column 12 of the table) except in instances of a head-on collision with $r_p=0$ (in this case, we assume $\Delta t = 2t_{\rm{orb}}$; column 8 of the table).

We show as different symbols the peak luminosity calculated for different values of the $s$ exponent in Equation~(\ref{eq:Lum}) which parameterizes the fraction of material transported from the outer edge of the disk to the black hole. We show $s=0$ (circles) as the most optimistic case, $s=1$ (stars) as the most pessimistic, and $s=0.5$ (triangles) as perhaps the most realistic case \citep[see, e.g.,][]{Yuan2012}. Different colors denote penetration depth of the interaction. As shown, less penetrating interactions (higher $r_p/r_T$; lighter colors) yield lower mass bound to the black hole and also lower accretion luminosities, as intuitively expected.

In the most optimistic case, our simulations suggest peak luminosities up to roughly $10^{49}\,\rm{erg/s}$ are feasible. This is comparable to the luminosity expected for long gamma ray bursts \citep[e.g.,][]{WandermanPiran2010,Petrosian2015}, depending of course on the uncertain beaming factor. Along these lines, the general outcome of these simulations --  a black hole embedded within a thick rotating torus of gas -- is qualitatively similar to a collapsar. Collapsars have been widely recognized as possible power sources for long gamma bursts \citep[e.g.,][]{MacFadyen2001}, which naturally prompts the question of whether these TDEs may lead to similar transients. Indeed, \citet{Perets_2016} pointed out that TDEs similar to those studied here may plausibly power ultra-long gamma ray bursts \citep[e.g.,][]{Levan2014}, whose exact origin remains poorly understood.
Detailed exploration of this possibility requires treatment of physical processes well beyond these hydrodynamic models. We reserve study of this for future more focused studies. However from the perspective of energy budget alone, a gamma ray burst outcome is not necessarily ruled out.

For practically all values of $s$, the accretion rate onto the black hole is reduced and the majority of the disk mass is lost in the form of a radiatively driven wind as shown by (magneto)hydrodynamic simulations of radiatively inefficient accretion flows \citep[e.g.,][]{Stone1999,Igumenshchev2000,Hawley2001,Narayan2012,Yuan2014}. In this case, the energy generated by accretion onto the black hole (which is reduced due to the reduced accretion rate) at the inner edge of disk will be reprocessed by this wind and released at the photon trapping radius \citep[e.g.,][]{LoebUlmer1997,StrubbeQuataert2009,MetzgerStone2016,Dai2018,Kremer2019c,PiroLu2020,Metzger2022}.
The emerging luminosity (which will be smaller than the accretion luminosity by a factor of roughly 100 due to adiabatic losses) is expected to be of order $10^{40}-10^{44}\,\rm{erg/s}$ and peak in the optical/UV band for temperatures expected here \citep[for more detail, see][]{Kremer2019c}.

\subsection{Multiple passages}

A key feature not discussed in previous studies is the occurrence of multiple disruptions during multiple pericenter passages after the star is tidally captured by the black hole. These multiple passages may have several observational consequences.

First, in many cases, the viscous accretion time of the disk formed after the first passage is less than the orbital period for the partially disrupted remnant to return (compare columns 12 and 13 of Table \ref{table:sims}). In this case, each pericenter passage may produce its own unique transient with properties roughly comparable to those described in Section \ref{sec:first_passage}. In general, our simulations show that the total mass stripped from the star (and therefore total mass bound to the black hole) increases with each passage (see Figure \ref{fig:evol_insp_all}). In particular, the typical \textit{final} outcome (after complete inspiral and full disruption of the star) is for roughly half of the initial stellar mass to be bound to the black hole. This suggests that successive accretion flares may exhibit a successive increase in brightness with the final flare being the brightest. Observations of repeated accretion flares (in X-ray from black hole accretion or in optical/UV from wind reprocessing) with these properties would hint strongly at this process. Indeed, observations such as the variable X-ray source HLX-1 \citep{Farrell2009} -- suggested to be an accreting stellar/intermediate-mass black hole in a globular cluster -- may already hint at this process \citep[e.g.,][]{Godet2014}.

An additional mechanism for producing a bright transient in the multiple passage case is shocks from disk wind ejecta. As summarized in Section \ref{sec:first_passage}, in the case of highly inefficient transport of material from the outer to inner edge of the disk such that the majority of disk mass is lost in a wind, the disk wind creates a quasi-spherical dense medium surrounding the black hole. In the event of a second pericenter passage, a second disk will form creating its own disk wind. As the second wind expands and collides and shocks with the material from the first wind, a fraction of its kinetic energy will be converted to radiation with luminosity

\begin{multline}
    \label{eq:Lwind}
    L_{s} \approx \frac{M_w v_w^2/2}{\max(P_{\rm{orb}}, t_{\rm diff})} \\
    \sim 3\times 10^{42} \mathrm{erg/s}\frac{M_w}{0.1 M_{\odot}} \Bigg( \frac{v_w}{0.01c} \Bigg)^2 \Bigg( \frac{P_{\rm{orb}}}{1\,\rm{mon}} \Bigg)^{-1},
\end{multline}
where $M_w v_w^2/2$ is the kinetic energy of the colliding wind shells, $P_{\rm{orb}}$ is the orbital period of the star to return to pericenter after the first passage (the delay time between launching of the first and second winds), and $t_{\rm diff}= \kappa M_w/(4\pi P_{\rm orb} v_w c) \simeq 2\mathrm{\,day}\, (M_w/0.1M_\odot) (v_w/0.01c)^{-1} (P_{\rm orb}/1\,\rm mon)^{-1}$ is the radiative diffusion time of the wind near the collision radius $R_w\sim P_{\rm orb}v_{\rm w}$ ($\kappa\simeq 0.3\rm\, cm^2\,g^{-1}$ being the Rosseland-mean opacity). For the fiducial parameters in the above equation, we have adopted a wind velocity $v_w\approx 0.01c$ from the analytic predictions in \citet{Kremer2019c} and assumed that the diffusion time is shorter than the orbital period. The optical depth near the collision radius, $\tau\approx \rho \kappa R_w \approx M_w \kappa/(v_w P_{\rm orb})^2\approx100$ (using same fiducial parameters as in Equation~\ref{eq:Lwind}), is much larger than unity. In this optically-thick regime, the color temperature of the emission is $T \simeq L_s/(\sigma_{\rm SB}4\pi R_w^2)\sim 10^4\mathrm{\,K}$ for our fiducial parameters, so we expect many of these shock-powered transients to be observable in the optical/UV bands.

\section{Event rates}
\label{sec:rates}

\begin{deluxetable*}{l|cc|c}
\tabletypesize{\footnotesize}
\tablewidth{0pt}
\tablecaption{Local universe rate estimates for various outcomes predicted from SPH simulations \label{table:rates}}
\tablehead{
	\colhead{Outcome} &
	\colhead{Stellar mass range} &
	\colhead{$\xi=r_p/r_T$ range} &
	\colhead{Rate [$\rm{Gpc}^{-3}\,\rm{yr}^{-1}$]}
}
\startdata
Full disruption & $0.1-100\,M_{\odot}$ & $0-0.4$ & 25 \\
\hline
Partial disruption+unbound remnant & $0.1-0.7\,M_{\odot}$ & $0.4-1.5$ & 44 \\
\hline
Partial disruption+tidal capture & $0.1-0.7\,M_{\odot}$ & $1.5-2$ & 22 \\
 & $0.7-100\,M_{\odot}$ & $0.4-2$ & 9 \\
\hline
\hline
Total & $0.1-100\,M_{\odot}$ & $0-2$ & 100 \\
\enddata
\tablecomments{\footnotesize Rough estimate of the local universe rate (from Equation \ref{eq:rate}) for the various outcomes described in Section \ref{sec:results} for relevant main-sequence star mass ranges and pericenter distance ranges. These relative rates are computed assuming a total TDE rate of roughly $100\,\rm{Gpc}^{-3}\,\rm{yr}^{-1}$ as estimated in \citet{Kremer2021a}.
}
\end{deluxetable*}

For a typical globular cluster with roughly 50-100 stellar-mass black holes at present \citep[e.g.,][]{Weatherford2020}, recent studies estimate a TDE rate of roughly $10^{-8}\rm{yr}^{-1}$ per cluster or, for a MW-like galaxy with roughly $\rm{a\,few}\times 10^2$ clusters, a rate of roughly 
$\rm{a\,few}\times 10^{-6}$ per galaxy. Assuming a globular cluster number density of roughly $1\,\rm{Mpc}^{-3}$ \citep[e.g.,][]{Rodriguez2015}, this corresponds to a TDE rate of roughly $10\,\rm{Gpc}^{-3}\,\rm{yr}^{-1}$ in the local universe, as shown in previous studies \citep[e.g.,][]{Perets_2016,Kremer2019c}.

As the properties (e.g., typical core densities) of old globular clusters are reasonably well known \citep[e.g.,][]{Harris1996} and the total number of black holes expected in old globular clusters are reasonably well constrained from numerical work \citep[e.g.,][]{Morscher2015,Askar2017,Weatherford2020,Kremer2020} and observations \citep[e.g.,][]{Strader2012,Giesers2019}, this $10\,\rm{Gpc}^{-3}\,\rm{yr}^{-1}$ value is reasonably robust (e.g., to within an order-of-magnitude). On the other hand, for other cluster types such as young star clusters \citep[YSCs; e.g.,][]{PortegiesZwart2010} and nuclear star clusters \citep[NSCs; e.g.,][]{Georgiev2016}, the total TDE rate is a bit more uncertain. In \citet{Kremer2021a}, we showed that these TDEs may occur in YSCs at rates ranging from roughly $2-200\,\rm{Gpc}^{-3}\,\rm{yr}^{-1}$ in the local universe, depending on the fraction of star formation expected to occur in star clusters and depending upon the uncertain details of the properties of YSCs at birth. \citet{Fragione2021} showed that stellar-mass black hole TDEs may occur in nuclear star clusters at rates of up to roughly $1-10\,\rm{Gpc}^{-3}\,\rm{yr}^{-1}$. Of course in massive NSCs with velocity dispersions of roughly $100\,\rm{km/s}$ or higher, the dynamics of TDEs may be quite different from those modeled in this study where we have assumed $v_\infty=0\,\rm{km/s}$ representative of lower-mass clusters.

In clusters where the gravitational-focusing regime applies (e.g., excluding nuclear star clusters), the cross section for such TDEs (and thus the rate, $\Gamma$), varies linearly with radius: $d \Gamma \propto dr$. In this case, close encounters where the black hole and star physically collide are just as likely as more distant encounters near the classical tidal disruption radius or beyond. Furthermore, as shown in \citet{Kremer2019c}, the distribution of main-sequence star masses that undergo TDE follows closely the initial stellar mass function giving us $d \Gamma \propto m^{-\alpha}dm$. In this case, for a stellar mass function ranging from roughly $0.1-100\,M_{\odot}$, the total rate scales as

\begin{equation}
    \Gamma \propto \int_{0.1}^{100} \Bigg(\int_0^{2r_T} dr\Bigg) m^{-\alpha}  \,dm.
\end{equation}
Assuming for simplicity a single component mass function with slope $\alpha=2.35$ \citep[e.g.,][]{Salpeter1955}\footnote{The true initial mass function of star clusters is uncertain. Some recent studies \citep[e.g.,][]{Marks2012} argue a more-top heavy mass function is likely for some clusters, especially at low metallicity. In this case the relative rates of TDEs across mass function may differ from those presented here.}, a main-sequence star mass-radius relation of $R\propto M^{0.6}$ (appropriate for $M\gtrsim 0.7\,M_{\odot}$ and a reasonable approximation for $0.1 \lesssim M\lesssim 0.7\,M_{\odot}$), and a fixed black hole mass of $10\,M_{\odot}$, we estimate the total rate of events with main-sequence masses in the range $m_1$ to $m_2$ and pericenter distances in the range $r_{p,1}=\xi_1 r_T$ to $r_{p,2}=\xi_2 r_T$ as\footnote{Note that the $\xi=r_p/r_T$ parameter defined here should not be confused with the traditional $\beta=r_T/r_p\gtrsim1$ parameter commonly used in the TDE literature to quantify the penetration depth of an encounter.}

\begin{multline}
    \label{eq:rate}
    \Gamma \approx 4.5 \,(\xi_2-\xi_1) \int_{m_1}^{m_2}m^{-2.08} \,dm  \,\rm{Gpc}^{-3}\,\rm{yr}^{-1}
\end{multline}
where we have additionally assumed the total rate integrated over all masses and pericenter distances is roughly $100\,\rm{Gpc}^{-3}\,\rm{yr}^{-1}$ as estimated in \citet{Kremer2021a}. Here $\xi_1$ and $\xi_2$ are the dimensionless penetration factors of the lower and upper bounds in pericenter distance. As shown in our SPH models, $\xi \in [0,2]$ captures the full parameter space of interest.

In Table \ref{table:rates}, we show the total rate estimated from this scaling for the various outcomes predicted in our SPH models. We use stellar mass and pericenter distance ranges appropriate for each outcome based on the results from the simulations (see in particular Figure \ref{fig:summary}). To within a small factor, we predict the three outcomes (full disruption, partial disruption+unbound remnant, and partial disruption+tidal capture) occur at roughly comparable rates in the local universe. Of course, these are very simple estimates intended to give a rough sense of the relative rates of the various outcomes identified in this study. Future studies may constrain these rates in more detail by (1) performing finer grids of SPH models to determine more precisely the boundaries between the various outcomes and (2) self-consistently computing the various possibilities within a direct N-body cluster simulation.

Finally, we note that the dynamically-active cores of dense star clusters may not be the only environment capable of facilitating such TDEs. For hierarchical triples (or, in principle, quadrupoles or higher multiples) with a black hole+main-sequence star inner binary, the Lidov-Kozai mechanism may drive eccentricity oscillations in the inner binary that may lead to tidal disruption \citep{Fragione2019}. We reserve for a future study an examination of how the TDE dynamics and rates may differ in this regime.

\section{Summary and Discussion}
\label{sec:summary}

\subsection{Summary}

We have performed a suite of SPH simulations of close encounters of main-sequence stars and stellar-mass black holes typical of those expected to occur in dense star clusters. Our main findings are as follows:

\begin{enumerate}
    \item For nearly head-on collisions ($r_p \sim 0$), the star is destroyed completely after a single passage. For more distant pericenter passages ($r_p\sim r_T$), the star is only partially disrupted by the black hole.
    \item For partial disruptions, the fate of the disrupted stellar remnant depends on its stellar structure. For stars with relatively uniform density distribution ($n=1.5$ polytropes -- e.g., low-mass M-dwarfs with fully or deeply convective envelopes), the large amount of material stripped from the star leads to an impulsive kick. In this case, the partially disrupted remnant remains unbound from the black hole and, in some cases, may be ejected from its host cluster.
    
    \item For more massive ($\gtrsim1\,M_{\odot}$) main-sequence stars, the partially disrupted star is tidally captured by the black hole. For the scenarios considered in this paper, the disrupted remnant returns to pericenter for one or more subsequent passages until ultimately being disrupted fully. The total number of pericenter passages depends upon the mass ratio and the penetration factor of the encounter.
    
    \item In all cases, a fraction of material stripped from the star becomes bound to the black hole and a fraction (with positive total energy) becomes unbound from the system entirely. This unbound ejecta causes a recoil kick to the center of mass of the final system (either a black hole+star binary or black hole+disk). If the ejecta mass is comparable to the black hole mass, the black hole may receive a significant kick ($v_{\rm{COM}}\gtrsim10\,$km/s). 
    
    \item Depending on the mass ratio and penetration factor, the total mass of the disrupted material bound to the black hole ranges from roughly $10^{-3}-\rm{a\,few}\,\it{M}_{\odot}$. With the exception of perfectly head-on collisions, the material bound to the black hole has non-zero angular momentum and thus settles into a rotating torus-like structure (a ``disk'') after roughly an orbital time (in general, $t_{\rm{orb}}$ is of order hours after the initial pericenter passage).
    
    \item For standard assumptions regarding the viscous accretion of thick disks, we predict accretion times of roughly days. Depending upon the uncertain details of the (super-Eddington) accretion process, a fraction of this material will accrete onto the black hole. For tidal disruptions of low-mass stars (the most common case; see Section \ref{sec:rates}), the mass bound to the black hole is typically in the range $0.05-0.5\,M_{\odot}$. For reasonable accretion physics assumptions ($s\sim0.5$ in Equation \ref{eq:Lum}), we expect a peak luminosity of the order $L_{\rm{peak}} \sim 10^{44 \pm 1}\,$ erg/s, which will mainly come out in the X-ray band. In principle, the energy budget may be comparable to that expected for long/ultra-long gamma-ray bursts \citep[e.g.,][]{Perets_2016}; however,  further work is required to test this possibility. If a fraction of disk material is ejected via a disk wind, this accretion luminosity is likely re-processed and larger radii and emitted as a bright optical/UV transient \citep[e.g.,][]{Kremer2019c}.
    
    \item When a partially-disrupted star is tidally captured by the black hole and undergoes additional passages, repeated accretion flares are possible. Detection of periodic transient sources (with repetitions likely occurring on timescales ranging from roughly days to years) would hint strongly at this mechanism. More detailed simulations that self-consistently include radiation/accretion feedback processes are necessary to test this outcome.
    
    \item Informed by our SPH results, we computed the relative event rate of the various predicted outcomes. To within a small factor, we predict full disruptions, partial disruptions with unbound remnants, and partial disruptions accompanied by tidal capture occur at roughly comparable rates in the local universe.
\end{enumerate}

\subsection{Future work}
\label{sec:future}

A key aspect not addressed in the SPH simulations presented here is the role of accretion feedback on the hydrodynamics. The binding energy of the disrupted debris bound to the black hole is roughly $E_{\rm{bind}} \approx GM_{\rm{BH}}M_{\rm{disk}}/R_{\rm{disk}}\approx 10^{49} \rm{erg}$ for $M_{\rm{BH}}\approx 10\,M_{\odot}$, $M_{\rm{disk}}\approx 0.3\,M_{\odot}$, and $R_{\rm{BH}}\approx R_{\odot}$.
Meanwhile, the energy released through accretion of $M_{\rm{acc}}$ onto the black hole can be estimated as $E_{\rm{acc}}\approx \epsilon M_{\rm{acc}}c^2$ (see Equation~\ref{eq:Lum}). For $\epsilon \approx 0.01$, only roughly $10^{-3}\,M_{\odot}$ of material must be accreted (a plausible value for $s\approx0.5$ in Equation~\ref{eq:Lum}) in order for $E_{\rm{acc}}$ to be comparable to $E_{\rm{bind}}$. Thus, the energy released through accretion may well unbind the remaining material bound to the black hole. Of course, this requires that the radiated accretion energy efficiently couple mechanically to the gas, which depends on the details of the accretion flow near the black hole (e.g., if jets are launched the emission could be narrowly beamed). Careful consideration of these details are beyond the scope of the present study, but this simple order-of-magnitude estimate shows that the role of feedback could, in principle, be important. Work is currently underway to incorporate accretion feedback as a ``sub-grid'' model into \texttt{StarSmasher} following similar principles to those employed in galaxy formation simulations \citep[e.g.,][]{Springel2005,Hopkins2015} with accretion feedback energy heating the surrounding gas.

In addition to the treatment of accretion feedback, several other elements remain unexplored here and may form the basis for future studies. For example, while our study employed simple polytropic stellar models, future calculations could start from more realistic stellar models computed with \texttt{MESA} \citep{Paxton2015}, following similar work using \texttt{StarSmasher} \citep[e.g.,][]{Hatfull2021}. This is especially important to model the tidal disruption of more evolved stars (e.g., giants) which have been proposed to play a role in the formation of compact black hole binaries like those observed in 47~Tuc and M10 \citep[e.g.,][]{Ivanova2017}. Second, while current models explore the parabolic encounter regime (representative of star clusters with low velocity dispersions), extension to hyperbolic encounters (specifically $v_\infty \gtrsim 100\,\rm{km/s}$) will be necessary to explore the outcomes of TDEs and collisions in more massive star clusters, including nuclear clusters  \citep{Fragione2021}. Third, our SPH calculations have focused on low-mass black holes ($\sim 10\,M_{\odot}$). However, if present in clusters, higher-mass black holes ($\gtrsim 10^2\,M_{\odot}$), i.e., intermediate-mass black holes \citep[IMBHs; for review, see][]{Greene2020}, will also undergo TDEs \citep[e.g.,][]{Rosswog2009,MacLeod2016}. A number of recent studies have demonstrated that IMBHs may form naturally in clusters through successive black hole mergers \citep[e.g.,][]{Rodriguez2019} or through stellar collisional runaways \citep[e.g.,][]{PortegiesZwart2004,Kremer2020b,Gonzalez2021}. Finally, as the database of SPH models across the full encounter parameter space grows, implementation of fitting formulae that capture the key effects (especially pertaining to black hole mass growth) into \textit{N}-body codes such as \texttt{CMC} will become possible. This will enable self-consistent treatment of the evolution of black hole masses over time in \texttt{CMC} through the effects of TDEs and collisions. This may have important implications for the overall black hole mass spectrum in star clusters and the mass spectrum of binary black hole mergers that are now detectable as GW sources \citep{Lopez2019,Rizzuto2021}.

\acknowledgments
We thank the anonymous referee for a careful review of the manuscript and many helpful comments. We thank Josh Fixelle for contributions in the early stages of this project. We also thank Byron Rich for helping to prepare workstations, purchased with a grant from the George I.\ Alden Trust, to run several of the simulations of this paper. KK is supported by an NSF Astronomy and Astrophysics Postdoctoral Fellowship under award AST-2001751. WL is supported by the Lyman Spitzer, Jr. Fellowship at Princeton University. This work was supported by NSF Grant AST-2108624 and NASA ATP Grant 80NSSC22K0722 at Northwestern University. 
This research was supported in part through the computational resources and staff contributions provided for the Quest high performance computing facility at Northwestern University, which is jointly supported by the Office of the Provost, the Office for Research, and Northwestern University Information Technology. 

\bibliographystyle{aasjournal}
\bibliography{mybib}

\end{document}